\documentclass[aps,pre,preprint,showpacs]{revtex4-1}
\usepackage{amssymb}
\usepackage{latexsym}
\usepackage{amsfonts}
\usepackage{graphicx}
\usepackage{float}
\usepackage{amsmath}
\usepackage{mathcomp}
\usepackage{epsfig}
\usepackage{hyperref}
\begin{document}
\title{Asymmetric coupling in two-lane simple exclusion process with Langmuir kinetics: phase diagrams and boundary layers}

\author{Arvind Kumar Gupta}
\email[]{akgupta@iitrpr.ac.in}
\author{Isha Dhiman}

\affiliation{Department of Mathematics, Indian Institute of Technology Ropar, Rupnagar-140001, India.}

\begin{abstract}
We study an open system composed of two parallel totally asymmetric simple exclusion processes with particle
attachment and detachment in the bulk. The particles are allowed to change their lane from lane-$A$ to lane-$B$, but not conversely. We investigate the steady-state behavior of the system using boundary layer analysis on continuum mean-field equations to provide the phase diagram. The structure of the phase diagram is quite complex and provides a complete insight about the steady-state dynamics. We examine two kinds of transitions in the phase plane: bulk transitions and surface transitions, qualitatively as well as quantitatively. The dynamics of shock formation, localization and their dependence on the system parameters and boundary rates have been investigated. We also justify the non-existence of downward shock in both the lanes using fixed point theory. Further, we examine the effect of increasing lane-changing rate on the steady-state dynamics and observe that the number of steady-state phases reduces with increase in lane-changing rate. Our theoretical results are supported with extensive Monte-Carlo simulation results.
\end{abstract}

\maketitle
\section{\label{intro}Introduction}
In nature, there exist a large number of stochastic systems, which attain non-equilibrium steady-state when driven by an external field. In contrast to the systems in equilibrium, the steady-state of non-equilibrium systems is characterized by a finite particle current and violates the detailed balance condition. A unified study of the general class of such systems, known as driven diffusive systems (DDS), is difficult due to the variety of non-equilibrium phenomena exhibited by them. Asymmetric simple exclusion process (ASEP) is the simplest exemplary model to study DDS, in which particles obey hard-core exclusion principle and hop in a preferred direction along the lattice. Despite their simplicity, these models are competent to efficiently explain
some complex non-equilibrium phenomena such as boundary-induced
phase transitions ~\cite{krug1991boundary, kolomeisky1998phase, popkov2007boundary}, phase separation
~\cite{krug2000phase}, spontaneous symmetry breaking ~\cite{clincy2001symmetry} and
 shock formation~\cite{evans2003shock, popkov2003localization, jiang2008weak} etc. Apart from being the central issue of academic interest, ASEP and its variants can successfully describe various physical, chemical and biological processes such as kinetics
of bio-polymerization ~\cite{macdonald1968kinetics}, protein synthesis ~\cite{shaw2003totally, chou2004clustered},
dynamics of motor proteins ~\cite{klumpp2003traffic}, gel electrophoresis ~\cite{widom1991repton}, vehicular traffic
~\cite{nagel1996particle} and modeling of ant-trails ~\cite{chowdhury2000statistical} etc.

In totally asymmetric simple exclusion process (TASEP) with finite lattice size, the boundaries of the system are connected to the particle reservoirs which maintain specific densities at both the ends. The non zero particle current in the system is preserved by these reservoirs only, whereas the total number of particles remains conserved in the bulk. Recently, a lot of attention has been given to the exclusion processes coupled with a bulk reservoir, where the particles can attach and/or detach at bulk sites (Langmuir kinetics (LK)). The additional attachment-detachment dynamics violate the particle conservation in the bulk.
If the non-conserving dynamics occur faster (slower) than a rate of $O(1/L)$, then LK (TASEP) dominates the steady-state. Thus, it is very crucial to rescale the time suitably to observe the interesting interplay between the particle conserving TASEP and particle non-conserving LK dynamics.

Single-channel TASEP coupled with LK have been studied comprehensively in the literature. An extension of TASEP, involving irreversible particle detachments from a single site in the lattice has been studied by Mirin and Kolomeisky ~\cite{mirin2003effect}. The important phenomenon of phase coexistence in single-channel TASEP with LK has been examined by Parameggiani \textit{et al.} ~\cite{parmeggiani2003phase}. Later, a detailed study about the competing dynamics of particle conservation (TASEP) and particle non-conservation (LK) in a single-channel lattice has been presented by Parameggiani \textit{et al.} ~\cite{parmeggiani2004totally}. One of the distinguishing features of single-channel TASEP coupled with LK observed is the localization of shocks in the bulk ~\cite{parmeggiani2003phase}. This is in contrast to the TASEP without LK, where shocks move with a constant velocity and are driven out of the system. The shock formation and its localization in single-channel TASEP with LK has also been investigated by Evans \textit{et al.} ~\cite{evans2003shock}. Similar characteristics of the shock have also been identified in system with interacting particles (KLS model) ~\cite{popkov2003localization, mukherji2007shocks}.

A large number of real processes such as vehicular traffic, motor protein dynamics and various systems of oppositely
moving particles ~\cite{nagel1996particle, howardsinauer, albertsmolecular} exist in physical world, which comprise
particles moving and shifting in more than one channel. In order to understand the dynamical aspect of particle non-conserving non-equilibrium systems more realistically, it becomes important to analyze multi-lane ASEP with LK. In this direction, Jiang \textit{et al.} ~\cite{jiang2007two} studied two-lane TASEP with particle creation and annihilation only in one of the two lanes. Moreover, the particles could jump from one lane to another with equal rates (symmetric coupling). They found that shocks in two lanes get synchronized when lane-changing rate crosses a threshold value. In the context of motor protein traffic, Wang \textit{et al.} ~\cite{wang2007effects} proposed a two-lane symmetrically coupled TASEP model with LK in both the lanes. The model has been investigated using mean-field approximation and it has been shown that the symmetry in lane-changing rates produces a finite-size jumping effect in the domain wall, which disappears with the increase in system size.

In spite of the substantial work done on single-lane as well as on two-lane TASEP with LK, the crucial case, where the particle exchange rates between the two lanes are unequal, has been neglected due to its complexities. The subject of this paper is to explore the consequences of asymmetric coupling conditions in a two-lane totally asymmetric simple exclusion process in the presence of Langmuir kinetics in both the lanes. It has been reported in the literature ~\cite{pronina2004two, pronina2006asymmetric, gupta2013coupling, shi2011strong} that asymmetric coupling in a two-channel TASEP without LK leads to more complex and significantly different structure of steady-state phase diagram from the one in symmetric coupling conditions. This motivates us to undertake the present study on an asymmetrically coupled two-lane TASEP with LK. Our model is inspired by the unidirectional motion of motor proteins along the protofilaments, their binding-unbinding and shifting to adjacent filaments ~\cite{howardsinauer, albertsmolecular}.

The paper is organized as follows. In the next section, we present the model, discuss its governing dynamical rules, obtain the continuum limit of the model equations using mean-field approximation and find their steady-state solution using boundary layer analysis. The phase diagrams and the density profiles obtained from theoretical results as well as Monte-Carlo simulations are discussed in Sec. \ref{PD}. Sec. \ref{SD} describes the dependence of various features of the domain wall on the system parameters. The effect of lane-changing rate on the dynamics is discussed in sec. \ref{effect}. In the concluding section, we summarize the results and possible extensions of our work.

\section{The model}
\label{model} The microscopic model consists of two parallel one-dimensional lattice channels each with $L$ sites, denoted by $A$ and $B$. We consider two-lane system with open boundaries in which the two ends are attached with particle reservoirs to maintain specific densities at the boundaries. The state of the system is characterized by a set of occupation numbers $n_{j}^{i}$ ($i=1,2,3,.....L$; $j= A,B$), each of which is either zero (vacant site) or one (occupied site). The system consists of indistinguishable particles distributed under the hard-core exclusion principle which ensures that any attempt made by a particle to jump to another site is successful only when the target site is empty. We impose the following dynamic rules. For each time step, a lattice site $(i,j)$ is randomly chosen. At entrance ($i=1$), a particle can enter the lattice with rate $\alpha$ when $n_{j}^{1}=0$ and at exit ($i=L$), particle can leave the lattice with rate $\beta$ when $n_{j}^{L}=1$. The following possibilities govern the particle hoppings in the bulk ($i=2,3,....L-1$) of lane-$A$.\\
Case (i) If $n_{A}^{i}=1$, then the particle will first try to detach itself from the site with a rate $w_d$. If it cannot detach from the $(i,A)$ site, it will jump to site $(i+1,A)$ with unit rate provided $n_{A}^{i+1}=0$; otherwise it shifts to lane-$B$ with rate $w$ if $n_{B}^{i}=0$.\\
Case (ii) If $n_{A}^{i}=0$, then a particle attach to the site with a rate $w_a$.

The dynamics in lane-$B$ are similar to those in lane-$A$ with the only exception that particles are forbidden to shift from lane-$B$ to lane-$A$.

The proposed two-channel model with open boundaries can be thought of as a particle non-conserving system comprising of two parallel TASEPs coupled with Langmuir kinetics. The mutual interaction between the two lattices is realized with admissibility of lane-changing process which is completely biased in one direction. It can also be seen as a generalization of one-channel TASEP coupled with LK to an asymmetrically coupled two-channel TASEP with LK.

The temporal evolution of bulk particle densities ($1<i<L$) in the two lanes can be computed from the following master equations.
\begin{equation}
\frac{d\textlangle n_{A}^{i}\textrangle}{dt}=\textlangle n_{A}^{i-1}(1-n_{A}^{i})\textrangle - \textlangle n_{A}^{i}(1-n_{A}^{i+1})\textrangle
+ \omega_a \textlangle 1-n_{A}^{i}\textrangle -\omega_d\textlangle n_{A}^{i}\textrangle
-\omega\textlangle n_{A}^{i}n_{A}^{i+1}(1-n_{B}^{i})\textrangle, \label{eq1}
\end{equation}
\begin{equation}
\frac{d\textlangle n_{B}^{i}\textrangle}{dt}=\textlangle n_{B}^{i-1}(1-n_{B}^{i})\textrangle - \textlangle n_{B}^{i}(1-n_{B}^{i+1})\textrangle
+ \omega_a \textlangle 1-n_{B}^{i}\textrangle -\omega_d\textlangle n_{B}^{i}\textrangle
+\omega\textlangle n_{A}^{i}n_{A}^{i+1}(1-n_{B}^{i})\textrangle, \label{eq2}
\end{equation}
where $\textlangle\cdots\textrangle$ denotes the statistical average.
At boundaries, the particle densities evolve as
\begin{eqnarray}
\frac{d\textlangle n_{j}^{1}\textrangle}{dt}=\alpha\textlangle (1-n_{j}^{1})\textrangle - \textlangle n_{j}^{1}(1-n_{j}^{2})\textrangle, \label{eq3}\\
\frac{d\textlangle n_{j}^{L}\textrangle}{dt}=\textlangle n_{j}^{L-1}(1-n_{j}^{L})\textrangle - \beta\textlangle n_{j}^{L}\textrangle. \label{eq4}
\end{eqnarray}
The mean-field approximation is imposed on the system by neglecting inter-particle correlations as
\begin{equation}
\textlangle n_{j}^{i}n_{j}^{i+1}\textrangle=\textlangle n_{j}^{i}\textrangle \textlangle n_{j}^{i+1}\textrangle. \label{eq5}
\end{equation}
The continuum limit of the model can be obtained by coarse-graining discrete lattice with lattice constant $\epsilon=1/L$ and rescaling the time as $t{'}=t/L$. When the non-conserving processes in the system occur at a comparatively lower (higher) rate than particle conserving processes, the system attains stationary state locally due to conservative dynamics only. Thus rescaling the time variable is reasonable to understand the engagement between particle conserving and non-conserving dynamics. In order to observe the competing interplay between boundary and bulk dynamics, we also rescale the attachment, detachment and lane-changing rates in such a way that the kinetic rates decrease simultaneously with increase in system size ~\cite{parmeggiani2004totally}. So, we parameterize the vertical transition rates as follows:
\begin{equation}
\Omega_a=\omega_a L, \Omega_d=\omega_d L, \Omega=\omega L \label{eq7}
\end{equation}
To get the continuum limit of the model, we replace binary discrete variables $n_{j}^{i}$ with continuous variables $\rho_{j}^{i}\in [0,1]$ and retain the terms up to second-order in Taylor's series expansion (for large system i.e. $L>>1$) as
\begin{equation}
\rho_{j}^{i\pm 1}=\rho_{j}^{i}\pm \frac{1}{L}\frac{\partial\rho_{j}^{i}}{\partial x}+ \frac{1}{2L^2}\frac{\partial^2\rho_{j}^{i}}{\partial x^2}+ O\bigg(\frac{1}{L^3}\bigg).\label{eq6}
\end{equation}
Now, we drop the superscript $i$, as both the lattices are free of any kind of spatial inhomogeneity. The state of the two-channel system is described by the average densities ($\rho_A$ and $\rho_B$) in two lanes, which are functions of time $t{'}$ and quasi-continuous space variable $x\in [0,1]$, as
\begin{eqnarray}
\frac{\partial \rho}{\partial t{'}}+ \frac{\partial J}{\partial x}=S, \label{eq8}
\end{eqnarray}
where $\rho= \begin{bmatrix} \rho_A\\ \rho_B \end{bmatrix}$ , $J= \begin{bmatrix} -\frac{\epsilon}{2}\frac{\partial \rho_A}{\partial x}+ \rho_A(1-\rho_A)\\ -\frac{\epsilon}{2}\frac{\partial \rho_B}{\partial x}+ \rho_B(1-\rho_B) \end{bmatrix}$  and $S= \begin{bmatrix} \Omega_a (1-\rho_A)-\Omega_d \rho_A-\Omega \rho^{2}_A (1-\rho_B) \\ \Omega_a (1-\rho_B)-\Omega_d \rho_B+\Omega \rho^{2}_A (1-\rho_B) \end{bmatrix}$.\\\\

Here, $S$ represents the non-conservative terms formed by combination of lane-changing transitions and Langmuir kinetics. The components of $J$, denoted by $J_j; j= A, B$, are the currents in the particle conservation situation in lane-$A$ and $B$, respectively. In the continuum limit ($\epsilon\rightarrow 0$), the average current in both the lanes is bounded ($J_j \leq 1/4$). However, this bound holds only if the density is a smooth function of position $x$. If there appears a discontinuity in the density over a crossover region of width of $O(\epsilon)$, then the first order derivative term in the average current can not be ignored and the relation $J_j \leq 1/4$ does not hold any longer.

The coupling term ($\Omega \rho^{2}_A (1-\rho_B)$) in the non-conservative part arises due to the biased lane-changing phenomenon. In addition to the attachment and detachment kinetics, the coupling term acts as a sink for lane-$A$ and source for lane-$B$. In the absence of coupling term, two-channel system converts into two independent TASEPs with LK whose phase diagram has been well-studied in literature ~\cite{parmeggiani2004totally, mukherji2006bulk}. So, it will be interesting to investigate that how emergence of the coupling term into the macroscopic system affects the steady-state dynamics.

Reformulating system \eqref{eq8} in the steady-state, we have
\begin{eqnarray}
\frac{\epsilon}{2}\frac{d^2 \rho_A}{d x^2}+(2\rho_A-1)\frac{d\rho_A}{dx}+\Omega_a (1-\rho_A)-\Omega_d \rho_A-\Omega \rho^{2}_A (1-\rho_B)=0, \label{eq13}
\end{eqnarray}
\begin{eqnarray}
\frac{\epsilon}{2}\frac{d^2 \rho_B}{dx^2}+(2\rho_B-1)\frac{d\rho_B}{dx}+\Omega_a (1-\rho_B)
-\Omega_d \rho_B+\Omega \rho^{2}_A (1-\rho_B)=0. \label{eq14}
\end{eqnarray}

The boundary conditions for coupled nonlinear system of Eqs. \eqref{eq13} and \eqref{eq14} are $\rho_A(0)=\rho_B(0)=\alpha$, $\rho_A(1)=\rho_B(1)=1-\beta=\gamma$ (say)

The leading order terms in the above system play the similar role as performed by the vanishing viscosity term (regularizing term) in the Burgers' equation and their omission makes the coupled system over determined. Retaining second order terms in the system ensures to generate a smooth solution fitting all the four boundary conditions. The shocks or boundary layers (if any) are formed over regions of width of $O(\epsilon=1/L)$, across which a sudden rise/fall in the density profile occurs while current remains constant. This constancy in current is due to the irrelevance of particle non-conserving dynamics in the narrow boundary layer or shock region.

To understand the steady-state behavior of our system, we hereby employ leading order boundary layer analysis on the continuum mean-field equations. Being a general scheme to solve the hydrodynamic equation in the thermodynamic limit ~\cite{cole1968perturbation}, this approach has been quite successful in explaining the complete rich phase diagram of single-channel TASEP with LK ~\cite{mukherji2006bulk}. Here, we examine the steady-state dynamics for the particular case of equal attachment and detachment rates ($ \Omega_a=\Omega_d$). Under this case, Eqs. \eqref{eq13} and \eqref{eq14} reduce to
\begin{eqnarray}
\frac{\epsilon}{2}\frac{d^2 \rho_A}{d x^2}+(2\rho_A-1)\bigg(\frac{d \rho_A}{d x}-\Omega_d\bigg)
-\Omega \rho^{2}_A (1-\rho_B)=0, \label{eq15}
\end{eqnarray}
\begin{eqnarray}
\frac{\epsilon}{2}\frac{d^2 \rho_B}{d x^2}+(2\rho_B-1)\bigg(\frac{d \rho_B}{dx}
-\Omega_d\bigg)
+\Omega \rho^{2}_A (1-\rho_B)=0. \label{eq16}
\end{eqnarray}
Here, the coupling term complexifies the two-channel system which restrains us from finding the explicit solutions for average densities in both the lanes. We have employed combination of analytical and numerical techniques to obtain global approximate steady-state solution. The global solution is found by computing bulk and boundary layer solutions separately and then matching these solutions suitably.

In the thermodynamic limit ($L>>1$), the contribution of the regularizing terms is negligible and the major part of the density profile is described by solution of system of first-order equations. The solution obtained after ignoring second-order terms is known as outer solution or bulk solution. The outer solution of the overdetermined system is unable to meet the boundary conditions at both the boundaries simultaneously. This generates the notion of left outer and right outer solutions. The solution satisfying left (right) boundary condition is known as left (right) outer solution. Since, density profile has to satisfy the boundary condition at other end also, the global solution can not be given by outer solution alone. To satisfy the boundary conditions, a crossover narrow regime from left to right solution is formed which gives rise to either a boundary layer or a shock in the density profile. This solution is known as inner solution and is found by ignoring the non-conservative terms in the second-order equations.

Now, we need to solve the system of first order coupled ordinary differential equations (in the limit $\epsilon\rightarrow 0$) to obtain the outer solution in both the lanes. Though the elimination of second order terms simplifies the system, still it cannot be solved analytically because of the coupling terms. Moreover, the system is over determined due to which it cannot fulfil the four boundary conditions simultaneously. These limitations suggest us to use a suitable numerical scheme to get approximate outer solution of the continuum mean-field equations ~\cite{jiang2007two}. The density profiles in steady-state have been obtained by keeping the time derivative terms in the system and capturing the solution after sufficiently long time, which ensures the occurrence of steady-state. Four different outer solutions viz. $(\rho_{Al},\rho_{Bl}), (\rho_{Al},\rho_{Br}), (\rho_{Ar},\rho_{Bl})$ and $(\rho_{Ar},\rho_{Br})$, depending upon which boundary condition is satisfied by the outer solution in both the lanes, have been obtained. Here, $l$ and $r$ denote left and right solution, respectively.

The dynamics of inner solution can be studied by proper rescaling of space variable. For this, we introduce a new variable $\widetilde{x}=\frac{x-x_d}{\epsilon}$, where $x_d$ is the position of boundary layer. This rescaling leads to elimination of the source and sink terms in the hydrodynamic equations which is well justified as particle non-conserving dynamics are irrelevant in regions of width of $O(\epsilon)$. In terms of new variable $\widetilde{x}$, the equations governing inner solution in the thermodynamic limit can be expressed in a concise form ($j=A, B$) as follows
\begin{equation}
\frac{1}{2}\frac{d^2 \rho_{j,in}}{d \widetilde{x}^2}+(2\rho_{j,in}-1)\frac{d \rho_{j,in}}{d\widetilde{x}}=0. \label{eq19}
\end{equation}
Integrating once with respect to $\widetilde{x}$, we have
\begin{equation}
\frac{d \rho_{j,in}}{d \widetilde{x}}=2(a_j+\rho_{j,in}-\rho_{j,in}^{2}).\label{eq20}
\end{equation}
Here, $a_j$ is the constant of integration and is computed from the matching condition of outer and inner solutions.

Suppose, boundary layer appears at right boundary ($x=1$) for lane-$j$, the matching condition requires
\begin{equation}
\rho_{j,in}(\widetilde x \rightarrow -\infty)=\rho_{j,out}(x=1)=\rho_{j,o} (say). \label{eq21}
\end{equation}
Here, $\rho_{j,o}$ is value of left outer solution in lane-$j$ at $x=1$. Clearly, $\rho_{j,o}$ is a function of system parameters $\Omega_d$ and $\Omega$. Importantly, one should not infer from here that the inner solutions in both the lanes are independent of non-conservative dynamics. Moreover, the inner solution in lane-$A$ is influenced by the inner solution in lane-$B$ and vice-versa, although this might not appear explicitly by looking at the uncoupled system (Eq. \eqref{eq19}). However, the lane-changing and attachment-detachment phenomena impart their effect in the inner solution through the matching conditions.

Eq. \eqref{eq21} gives $a_j=\rho_{j,o}^{2}-\rho_{j,o}$, which physically interprets that current across the inner solution region must be equal to the bulk current entering the region. Solving Eq. \eqref{eq20} after substituting value of $a_j$, we can obtain the inner solution in lane-$j$ given by
\begin{equation}
\rho_{j,in}=\frac{1}{2}+\frac{|2\rho_{j,o}-1|}{2}\tanh\bigg(\frac{\widetilde x}{w_j}+\xi_j\bigg), \label{eq22}
\end{equation}
where $w_j=\frac{1}{|2\rho_{j,o}-1|}$ and $\xi_j=\tanh^{-1}\bigg(\frac{2\gamma-1}{|2\rho_{j,o}-1|}\bigg)$. The value of $\xi_j$ has been computed from the condition $\rho_{j,in}(\widetilde x = 0)=\gamma$. The left outer solution $\rho_{j,o}$ is a function of entrance rate $\alpha$ as it respects left boundary condition. So, $\xi_j$ becomes a function of $\alpha$ as well as $\gamma$. This dependence of inner solution on boundary rates gives rise to a region in $\alpha-\gamma$ phase-plane in which we get right boundary layer in lane-$j$ with positive slope and this region is a subregion of low-density (LD) phase and exists for $\gamma > \rho_{j,o}(\alpha)$. The solution given by Eq. \eqref{eq22} is referred to as $\tanh-r$ solution.

As $\widetilde x \rightarrow \infty$, the boundary layer at $x=1$ saturates to $\rho_{j,s}$ (say). The saturation of boundary layer is
mathematically expressed by the condition $\rho_{j,o}^{2}-\rho_{j,o}+\rho_{j,s}-\rho_{j,s}^{2}=0$ which gives $\rho_{j,s}=1-\rho_{j,o}$.
When $\gamma>\rho_{j,s}(\alpha)$, the inner solution fails to satisfy the right boundary condition
$\rho_{j,in}(\widetilde x\rightarrow \infty)=\gamma$ and deconfines from the boundary to enter the bulk of
lane-${j}$ in the form of a shock. Thus $\gamma=1-\rho_{j,o}(\alpha)$ acts as a bulk transition line between LD
and shock phases. Such kind of continuous transition is reminiscent of bulk transition observed in single-channel TASEP with LK ~\cite{mukherji2006bulk, mukherji2005nonequilibrium}, known as shockening transition.
The shockening transition shows power-law behavior with universal critical exponents ~\cite{mukherji2005nonequilibrium}.

Within LD phase, the slope of boundary layer is negative for $\gamma < \rho_{j,o}(\alpha)$ and the inner solution in this region is
\begin{equation}
\rho_{j,in}=\frac{1}{2}+\frac{|2\rho_{j,o}-1|}{2}\coth\bigg(\frac{\widetilde x}{w_j}+\hat{\xi}_j\bigg), \label{eq23}
\end{equation}
where $\hat{\xi}_j= \coth^{-1}\bigg(\frac{2\gamma-1}{|2\rho_{j,o}-1|}\bigg)$. The change in the slope of boundary layer describes a surface transition, which does not affect bulk density profile. One can predict that for every $\alpha$, a surface transition at $\gamma= \rho_{j,o}(\alpha)$ accompanies the bulk transition at $\gamma=1-\rho_{j,o}(\alpha)$. This leads to formation of two subregions inside the LD phase,  $\gamma< \rho_{j,o}(\alpha)$ and  $\rho_{j,o}(\alpha) < \gamma < 1 - \rho_{j,o}(\alpha)$, in which the density profile possesses a right boundary layer with positive and negative slope, respectively. The length scale
described by $\xi_j$ shows a logarithmic divergence ($\xi_j\thicksim \ln |\gamma- \rho_{j,o}|$)
as one approaches the surface transition line from either of the two subregions.

Now, we explain the surface transition from the physics point of view of the system. On the surface transition curve, the value of left outer solution at $x=1$ i.e. $\rho_{j,o}$ exactly matches the density at right boundary ($\gamma$). This generates a density profile with no boundary layer. Fixing $\alpha$, if one reduces withdrawal rate of particles (equivalently increases $\gamma$), particles start accumulating near right boundary forming an increasing boundary layer. Similarly, reducing $\gamma$ increases the withdrawal rate which creates scarcity of particles near right boundary and justifies the formation of a decaying right boundary layer. On the similar lines, we can analyze the boundary layer at $x=0$.
\section{\label{PD}Phase diagrams and density profiles}
With an aim to examine the effect of biased lane-changing phenomenon on the steady-state dynamics,
we extend our discussion in this section to draw the phase diagram of the
two-channel system for fixed values of $\Omega_d$ and $\Omega$.

Note that the steady-state dynamics of symmetrically coupled two-channel TASEP with LK are
similar to those in single-channel TASEP with LK (ignoring finite-size effects) ~\cite{wang2007effects}. The symmetry in coupling rates leads to the cancelation of lane-changing source terms with sink terms in the mean-field hydrodynamic equations, which gives two uncoupled ordinary differential equations representing two independent TASEPs with LK. Hence, the topology of phase diagram of single-channel TASEP with LK model remains preserved. This is totally in contrast to the asymmetric
coupling conditions, where we have existence of a novel phase diagram, considerably different from that of
single-channel TASEP with LK system ~\cite{mukherji2006bulk, parmeggiani2004totally}.

\begin{figure}
 \begin{center}
 \includegraphics[height=7.5cm,width=10cm]{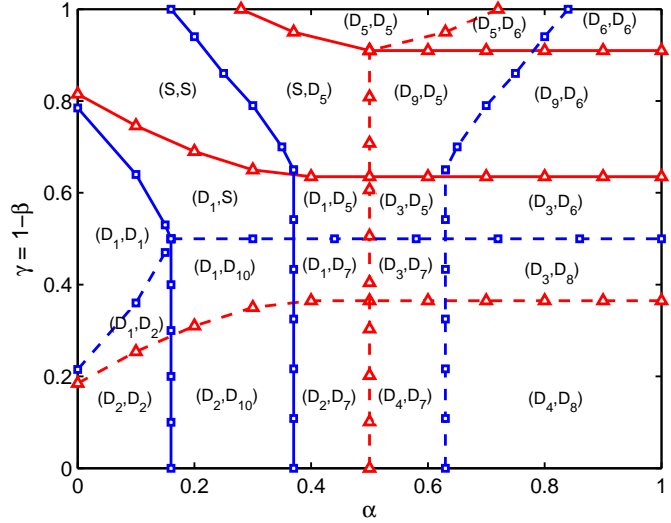}
 \caption{ \label{fig1}(Color online) Phase diagram of two-channel system for $\Omega_d=0.2$ and $\Omega=1$. $D_1$: tanh-r, $D_2$: coth-r, $D_3$: tanh-r with lbl, $D_4$: coth-r with lbl, $D_5$: tanh-l, $D_6$: coth-l, $D_7$: tanh-l with rbl, $D_8$: coth-l with rbl, $D_9$: S+lbl and $D_{10}$: S+rbl.  Here, S denotes shock, lbl and rbl denote left boundary layer and right boundary layer, respectively. Curves marked with triangles and squares represent phase boundaries of lane-$A$ and $B$, respectively. Solid and dashed curves denote bulk phase transitions and surface transitions, respectively.}
  \end{center}
\end{figure}

We have obtained the phase diagram for $\Omega_d=0.2$ and $\Omega=1$.
The biased lane-changing rule generates a quite richer and complex phase diagram (Fig.~\ref{fig1}).
The $\alpha-\gamma$ phase-plane is segmented into twenty-two distinct regions, each of which describes a different density profile. The curves marked with triangles (squares) denote phase change in lane-$A$ (B). The phase transitions in the bulk and surface transitions are shown with the help of solid and dashed curves, respectively.

\begin{figure}[htb] \begin{center}
\includegraphics[height=6cm,width=8cm]{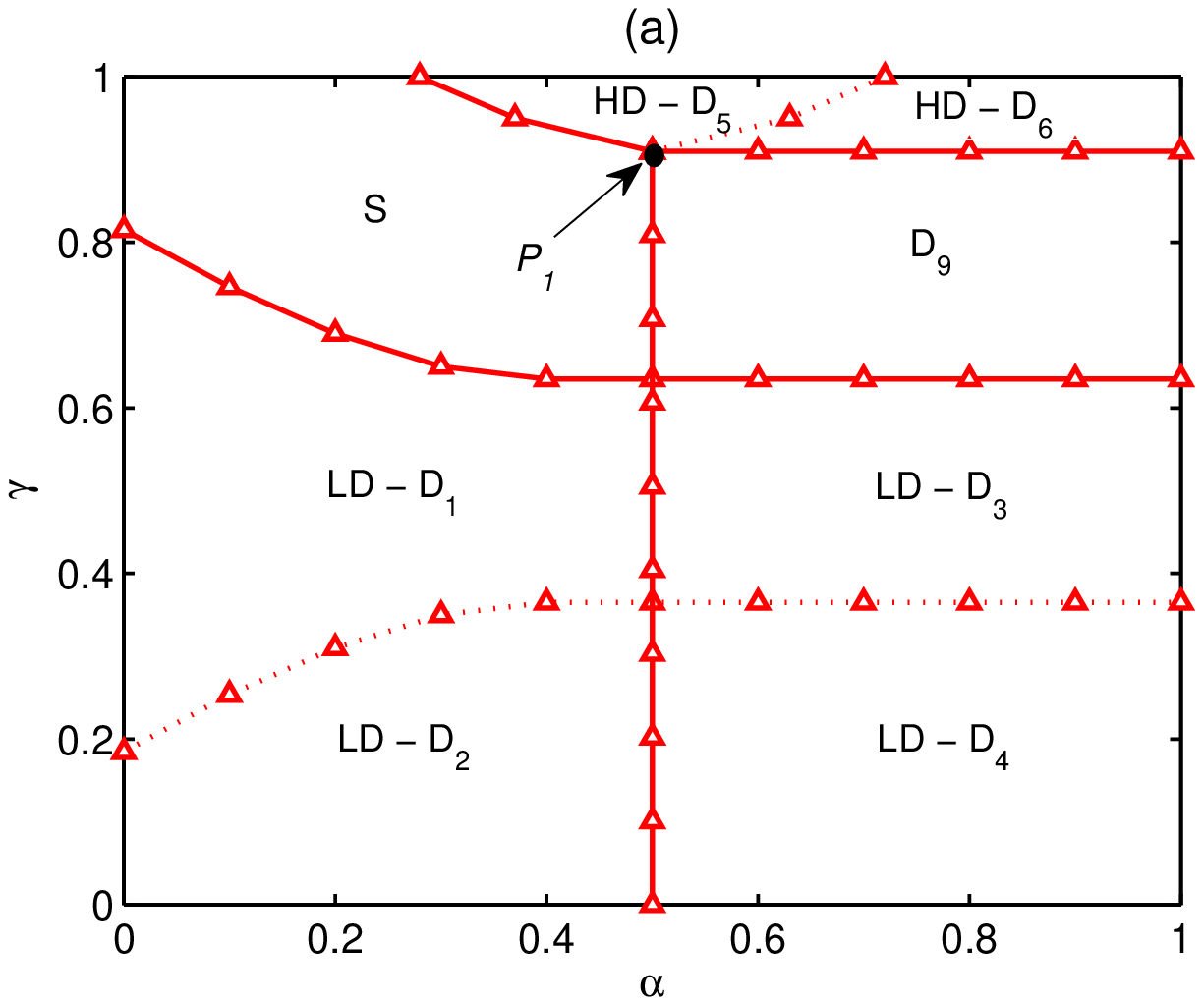}
\includegraphics[height=6cm,width=8cm]{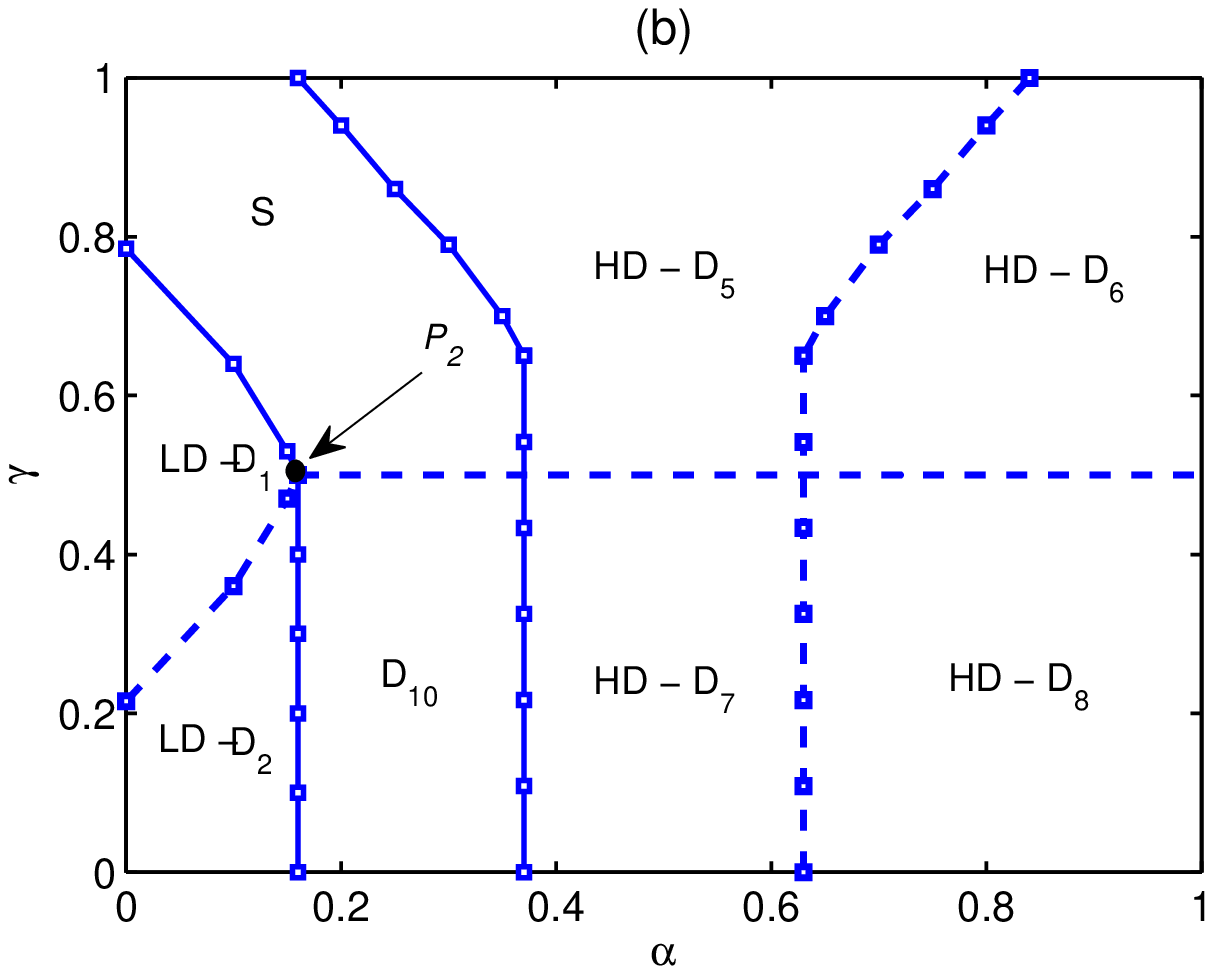}
 \caption{\label{fig2}(Color online) Phase boundaries in lane-$A$ and $B$ shown in Fig. $2(a)$ and Fig. $2(b)$, respectively. Indicated- LD: Low-density, HD: high-density and S: shock. The rest of the notations are same as used in Fig.~\ref{fig1}. $P_1$ and $P_2$ are the critical points and their origin is discussed in section \ref{PD}}.
\end{center}
\end{figure}

In order to gain deeper insight about the various transitions from one phase to another more appropriately, it is suitable to inspect the topology of the phases for each lane separately (Fig.~\ref{fig2}).
Both the lanes exhibit three distinct steady-state phases low-density (LD), shock (S) and high-density (HD).
The LD phase in $\alpha-\gamma$ phase-plane for lane-$A$ comprises of two major parts ($\alpha < 1/2$ and $\alpha > 1/2$), each of which is further divided into two subregions by a surface transition line. The surface transition corresponds to change in sign of slope of the right boundary layer. In $\alpha < 1/2$ region,
we find a boundary layer at $x=1$,whose properties have already been discussed in the previous section. When $\alpha> 1/2$, a decaying boundary layer at $x=0$ starts developing and grows in size with increase in the entrance rate. The appearance of the left boundary layer can be understood as follows. In LD phase, the bulk density is less than $1/2$, which is not compatible with the boundary condition $\rho_A(x=0)=\alpha (> 1/2)$. Thus, in order to satisfy the left boundary condition, a decaying boundary layer evolves at $x=0$.

The saturation of right boundary layer occurs at phase boundary between LD and shock phase. On further
increasing $\gamma$, we see the formation of a shock through shockening transition,
whereas dynamics at left boundary remains preserved. The line $\alpha=1/2$ divides the shock phase into two subregions. The density profile comprise of shock with no boundary layer and shock accompanied by a decaying left boundary layer for $\alpha< 1/2$ and $\alpha > 1/2$, respectively. For fixed value of $\alpha (< 1/2)$, further increase in $\gamma$ leads to leftwards motion of shock in the bulk until it reaches $x=0$ to produce HD profile with a tanh-type left boundary layer (tanh-l). The HD phase also involves change in sign of slope of the left boundary layer across the surface transition line (dashed). The two subregions formed by surface transition line are $1-\rho_{A,o}(\gamma) < \alpha < \rho_{A,o}(\gamma)$ and $\alpha > \rho_{A,o}(\gamma)$. Here, $\rho_{A,o}(\gamma)$ is the value of right outer solution at $x=0$. The intersection of lines $\alpha = \rho_{A,o}(\gamma)$ and $\alpha = 1- \rho_{A,o}(\gamma)$ locates a critical point $(\alpha_{cA},\gamma_{cA})$ in the phase-plane ($P_1$), where $\alpha_{cA}=1/2$ and the value of $\gamma_{cA}$ can be computed from $\rho_{A,o}(\gamma)= 1/2$. The phase boundary between shock and HD phase is given by
\begin{equation}
\alpha = 1-\rho_{A,o}(\gamma); \alpha < \alpha_{cA}
\end{equation}
Across this boundary, shock is formed due to deconfinement of the tanh-l type inner solution, which is similar to the shockening transition from LD to shock phase. On the other hand, the coth-l boundary layer in $\alpha > \rho_{A,o}(\gamma)$ does not deconfine to produce shock and $\gamma_{cA}$ remains the critical value of $\gamma$ for $\alpha > \alpha_{cA}$. This gives horizontal transition line $\gamma=\gamma_{cA}$ as the phase boundary between HD and shock phases.

The various phase boundaries in lane-$B$ can also be obtained in a similar fashion as discussed for lane-$A$. A large part of $\alpha-\gamma$ phase-plane for lane-$B$ is covered by HD phase which consists of two major subparts $\gamma< 1/2$ and $\gamma > 1/2$. There exist only left boundary layer of either $\tanh$ or $\coth$ type in $\gamma > 1/2$ subregion while for $\gamma < 1/2$, the density profile also incurs a decaying right boundary layer in order to meet
the right boundary condition. The deconfinement of left boundary layer in HD phase leads to shock formation. Similarly, the transition from LD to shock phase is also because of the deconfinement of tanh-r boundary layer for $\gamma > 1/2$, but coth-r boundary layer can not produce shock through deconfinement for $\gamma < 1/2$. This observation is similar as seen in HD phase for lane-$A$. The two lines $\gamma= \rho_{j,o}(\alpha)$ and $\gamma= 1- \rho_{j,o}(\alpha)$ intersect and gives a critical point
($P_2$) $(\alpha_{cB},\gamma_{cB})$, where $\gamma_{cB}=1/2$ and $\alpha_{cB}$ can be computed from
$\rho_{j,o}(\alpha)=1/2$. The vertical line $\alpha=\alpha_{cB}$ gives the transition line from LD to shock phase
for $\gamma < 1/2$.

Now, the fundamental question of address here is that how introduction of a small asymmetry in lane-changing rates
produces significant changes in the phase diagram. This can be understood on a more physical ground. Since,
we have allowed lane-changing only from lane-$A$ to $B$, the lane-$A$ can be thought of
as a homogeneous bulk reservoir of particles for lane-$B$. So, in addition to attachment and detachment occurring due to LK, more particles are getting detached from lane-$A$ and attached to lane-$B$, due to the biased lane-changing rules. This creates an imbalance between attachment and detachment rates in both the lanes. Now, effective detachment rate in lane-$A(B)$ has become higher (lower) than effective attachment rate in lane-$A(B)$. Due to this reason, the structure of phase diagram for lane-$A$ and lane-$B$ comes out qualitatively similar as that of single-channel TASEP with LK for $\Omega_a < \Omega_d$ (more detachment) and $\Omega_a > \Omega_d$ (more attachment), respectively ~\cite{mukherji2006bulk}. An interesting observation about the topology of phase diagrams of the two lanes is that a qualitative similar structure of one lane can be obtained by rotating the phase diagram of other lane by a right angle and then taking the mirror image with respect to vertical axis.

\begin{figure}[h]
\begin{center}
\includegraphics[height=6cm,width=8cm]{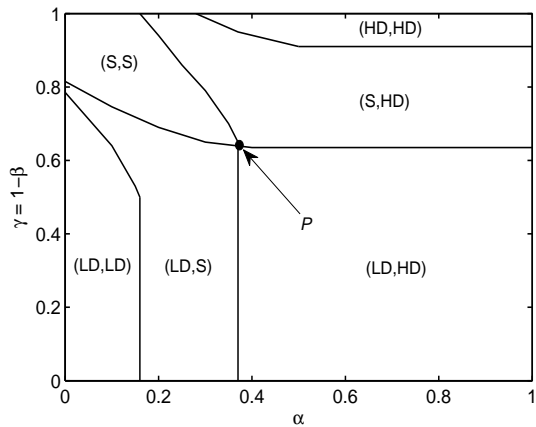}
\caption{Phase diagram for $\Omega_d=0.2$ and $\Omega=1$ showing only bulk phase transitions. The point formed by the intersection of the four different phase boundaries is marked $P$ and discussed in the text.}
\label{fig3}
\end{center}
\end{figure}

The complete picture about the steady-state phases (without surface transitions) is presented in Fig.~\ref{fig3}. There exists six distinct steady-state phases viz. (LD,LD), (LD,S), (S,HD), (S,S), (HD,HD) and (LD,HD). A qualitative comparison of Fig.~\ref{fig3} with the phase diagram of uncoupled system for $\Omega_a =\Omega_d$ ~\cite{mukherji2006bulk} reveals that a major part of the phase-plane is covered by (LD,HD) phase due to the absence of maximal-current (MC) phase in our system. Also, the region of LD(HD) phase expands while that of HD(LD) phase shrinks in lane-$A(B)$ by allowing biased lane-changing dynamics into the system. This is physically justified as shifting of additional particles from lane-$A$ to $B$ creates relative shortage of particles in lane-$A$ and abundance of particles in lane-$B$. Therefore, the average density in lane-$A$ is always lower than average density in lane-$B$ by virtue of which we have non-existence of (S,LD), (HD,S) and (HD,LD) phases in steady-state. The relation $\rho_A < \rho_B$ holds for all the phases irrespective of whether the two lanes exist in same or different phase.

\begin{figure*}[htb]
\begin{center}
\includegraphics[height=4cm,width=5cm]{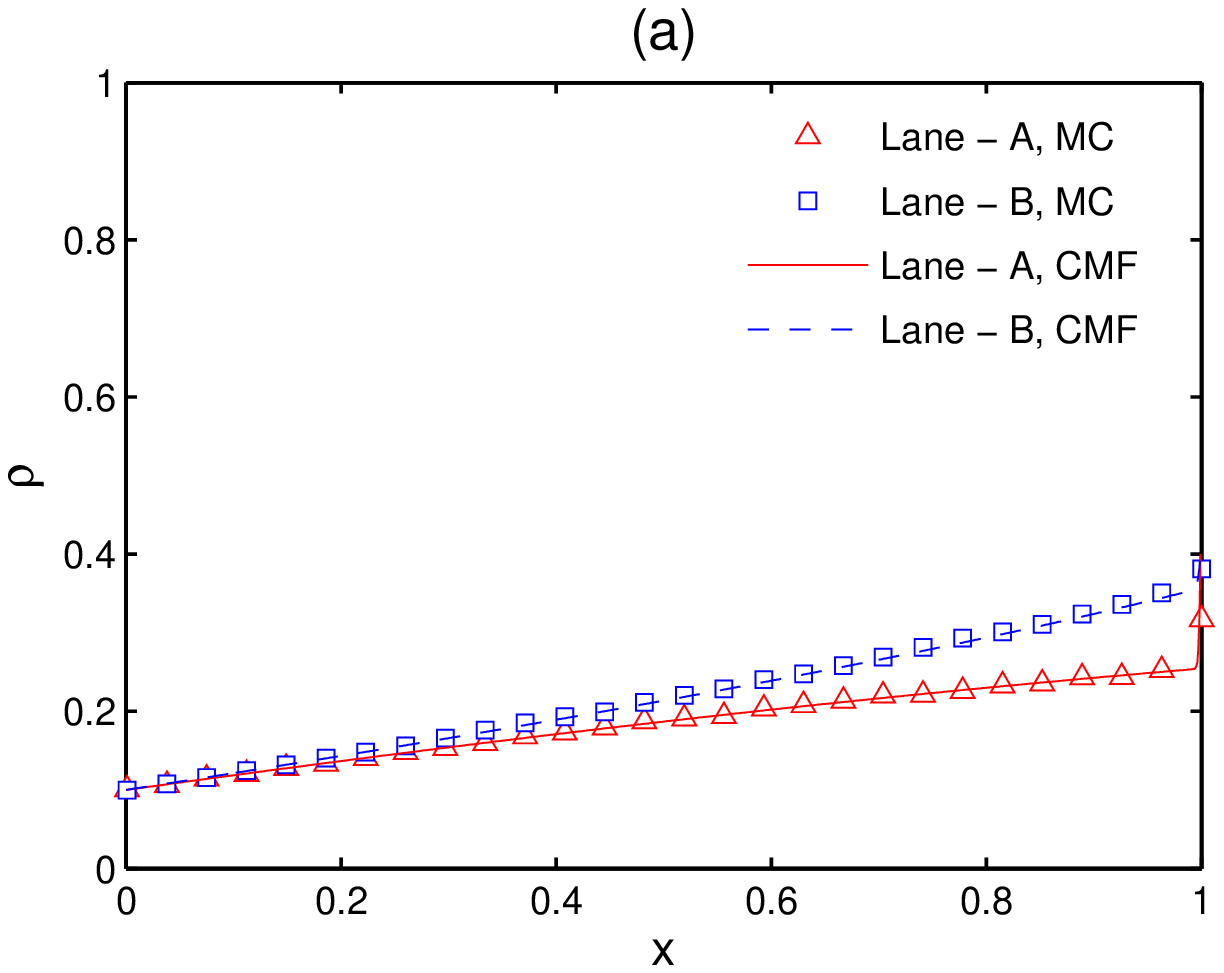}
\includegraphics[height=4cm,width=5cm]{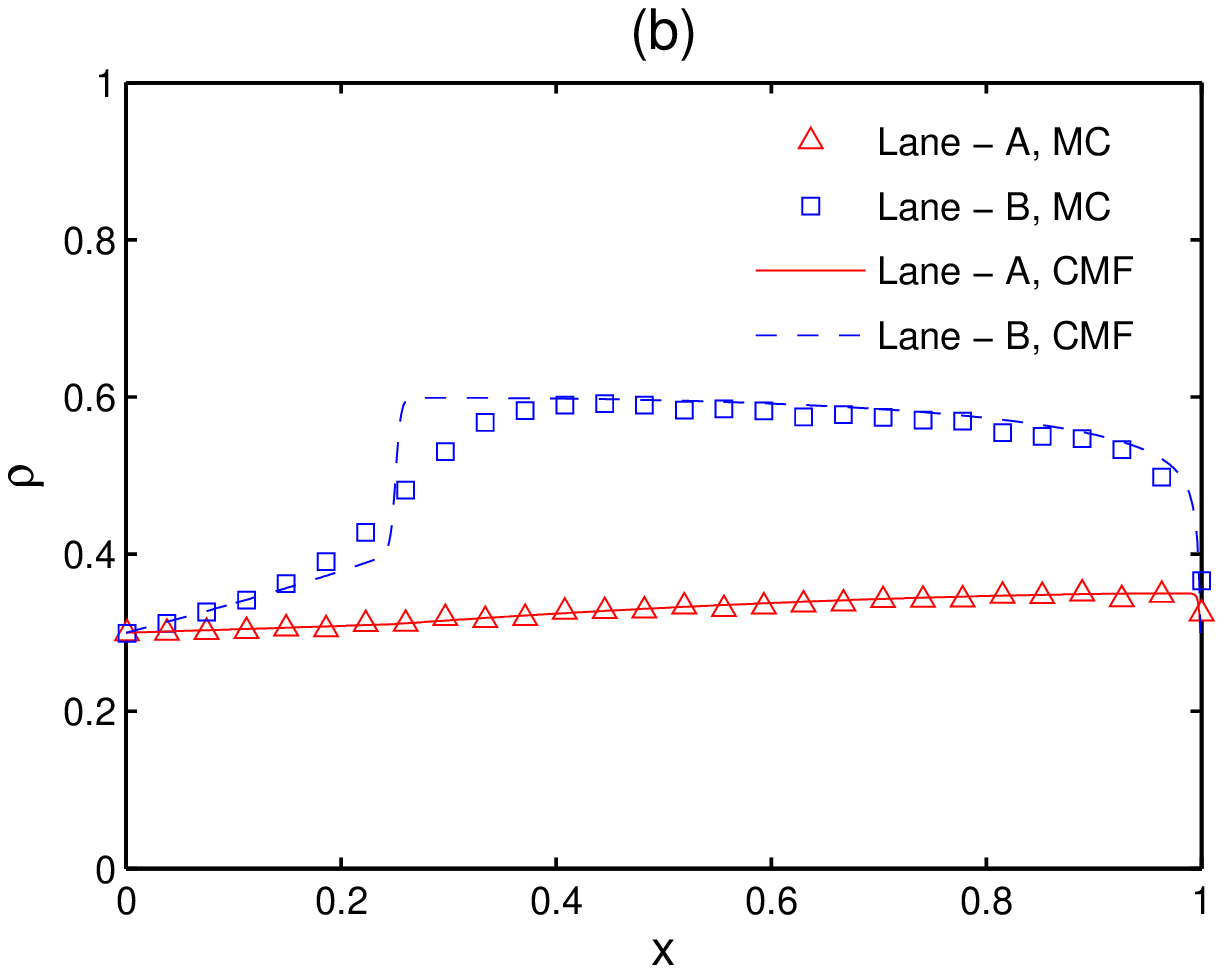}
\includegraphics[height=4cm,width=5cm]{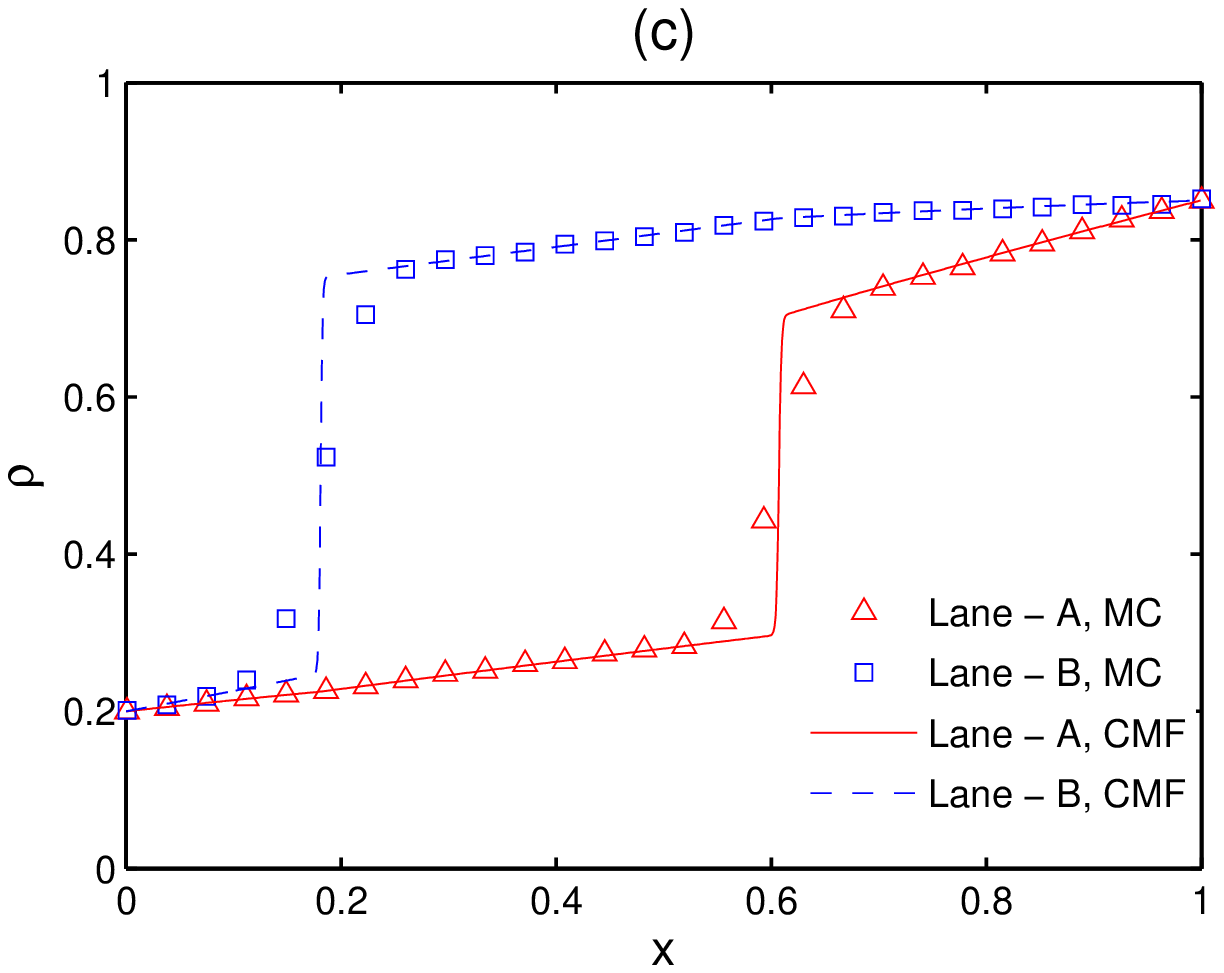}
\includegraphics[height=4cm,width=5cm]{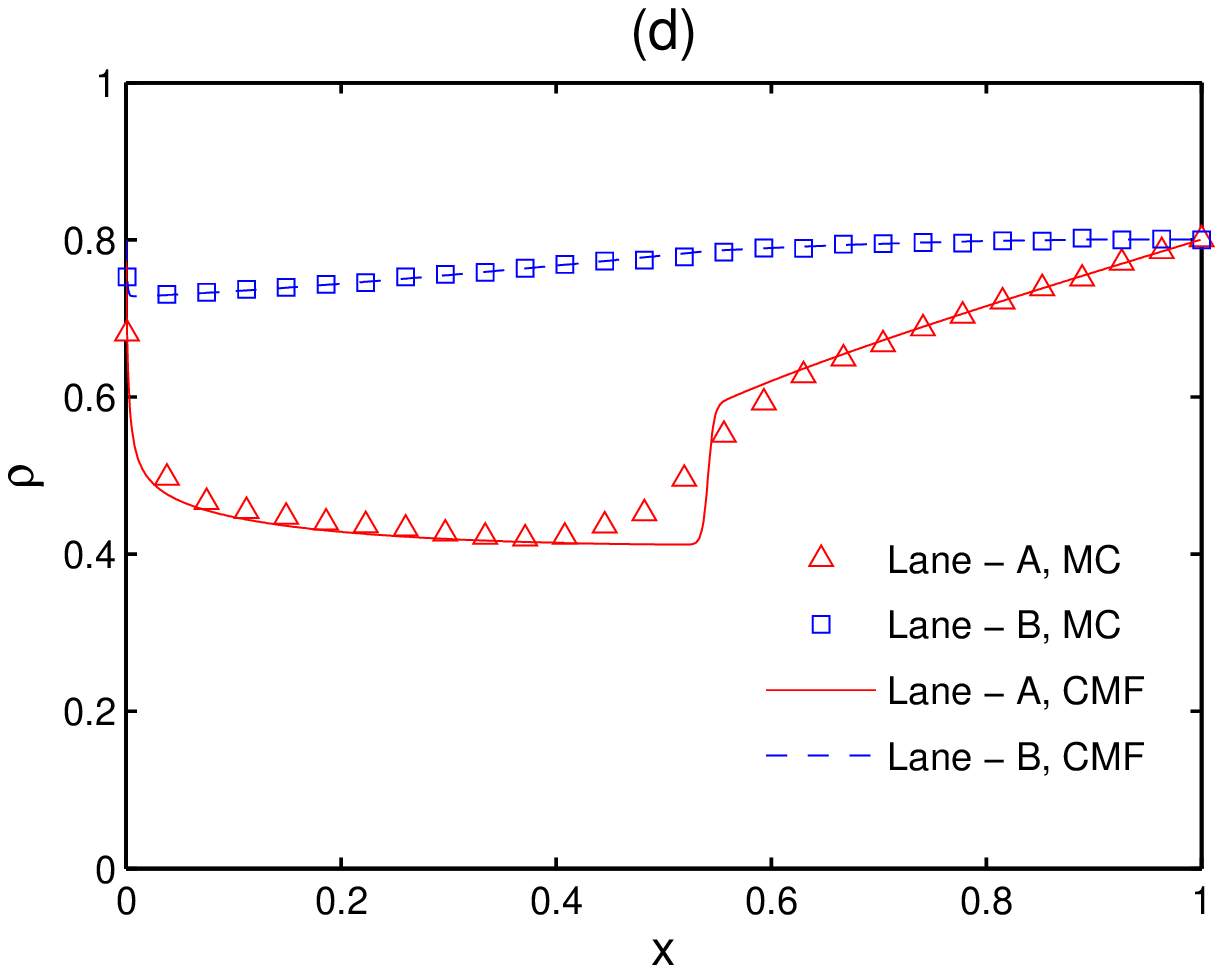}
\includegraphics[height=4cm,width=5cm]{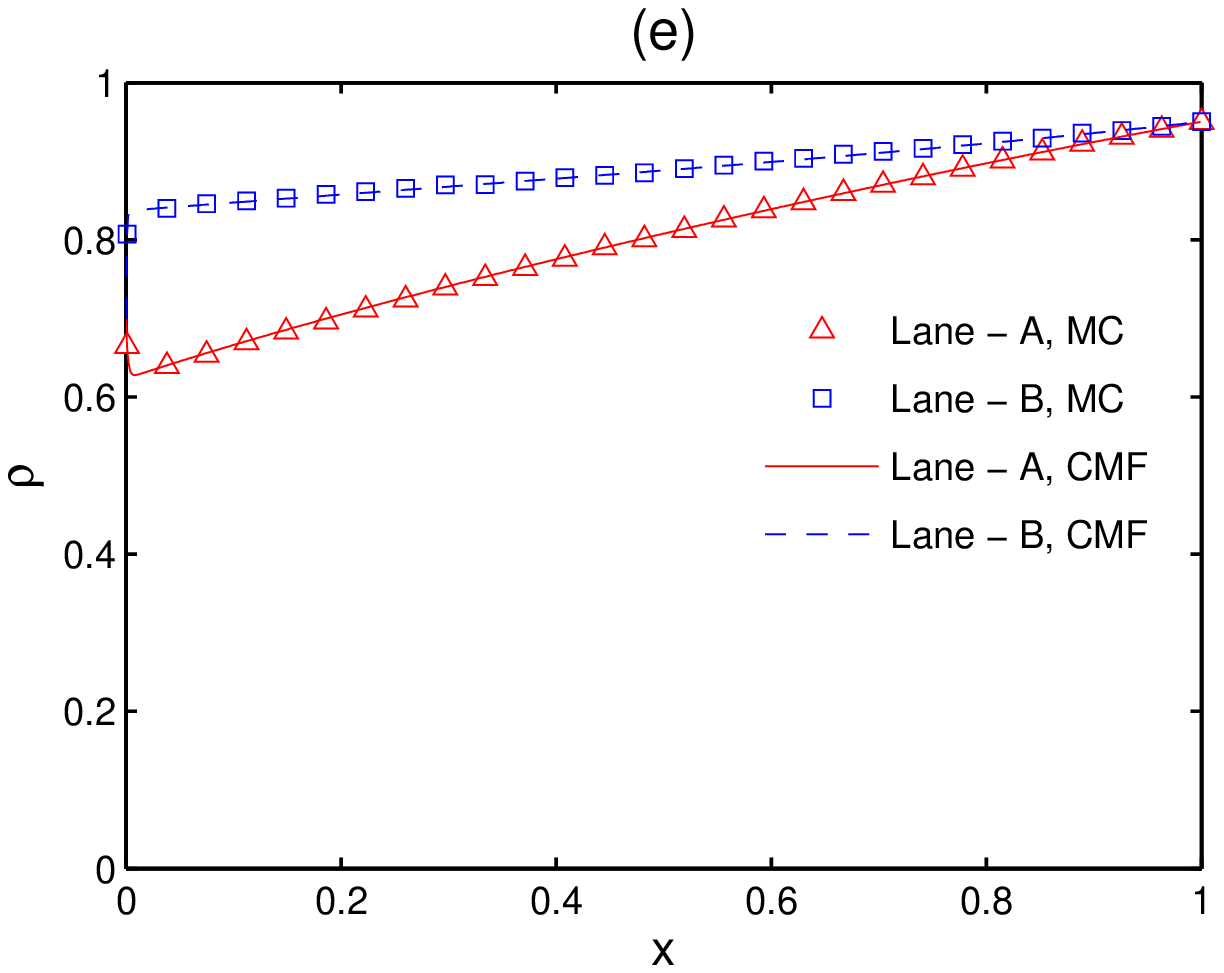}
\includegraphics[height=4cm,width=5cm]{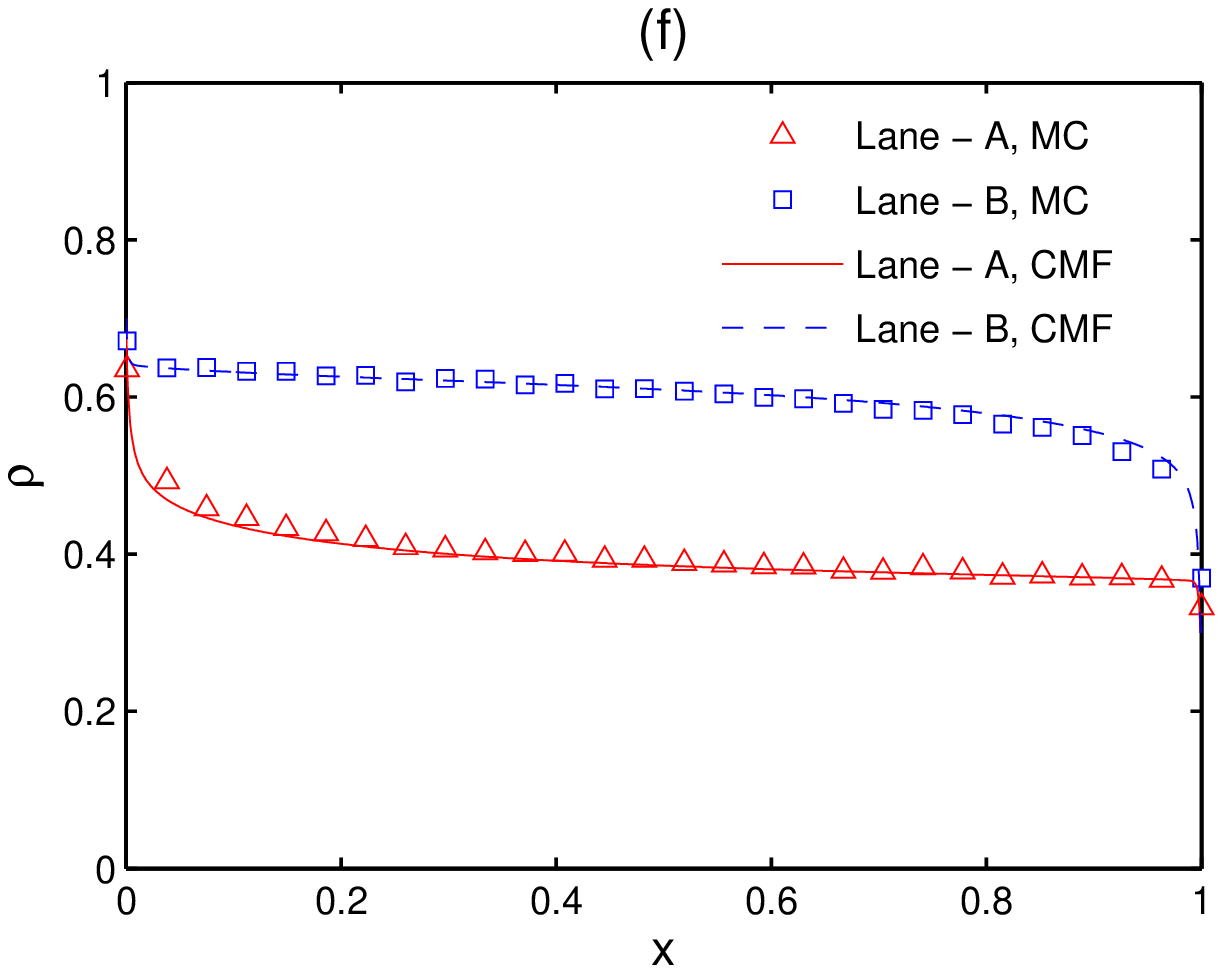}
\caption{\label{fig4} (Color online) Density profiles in (a) (LD,LD) phase for $\alpha=0.1, \gamma=0.4$, (b) (LD,S) phase for $\alpha=0.3, \gamma=0.3$, (c) (S,S) phase for $\alpha=0.2, \gamma=0.85$, (d) (S,HD) phase for $\alpha=0.8, \gamma=0.8$, (e) (HD,HD) phase for $\alpha=0.7, \gamma=0.95$ and (f) (LD,HD) phase for $\alpha=0.7, \gamma=0.3$. The red (solid) and blue (dashed) curves represent continuum mean-field density profiles in lane-$A$ and lane-$B$, respectively. The curves marked with triangles and squares show Monte-Carlo simulation results for lane-$A$ and $B$, respectively. The Monte-Carlo simulation results have been carried out for $10^9-10^{10}$ time steps for $L=1000$ after ignoring first $5\%$ time steps and are found to well agree with continuum mean-field results. }
\end{center}
\end{figure*}

In (LD,LD) phase, there exist right boundary layer of tanh or coth type in both the lanes (Fig.~\ref{fig4}(a)). The bulk transition lines $\alpha=\alpha_{cB}$ (for $\gamma\leq 1/2$) and $\gamma=\rho_{B,o}(\alpha)$ (for $\gamma > 1/2$) form the phase boundaries of (LD,LD) phase, crossing which one enters (LD,S) phase. In (LD,S) phase, lane-$A$ continues to be in LD state while in lane-$B$, a shock connecting left and right outer solutions emerges. For $\gamma < 1/2$, along with the shock, a decaying boundary layer is observed at right boundary as well (Fig.~\ref{fig4} (b)).
While keeping $\alpha$ fixed, we find a bulk phase transition in the two-channel system from (LD,S) phase to (S,S) phase on increasing $\gamma$. In this phase, shock is present in both the lanes (Fig.~\ref{fig4}(c)).
The curves $\gamma=1-\rho_{A,o}(\alpha)$ and $\alpha=1-\rho_{B,o}(\gamma)$ represent the boundaries of (S,S) phase and intersect at a critical point ($\alpha_c,\gamma_c$), $P$ in phase-plane. The shock in lane-$B$ moves leftwards in the bulk with increase in $\alpha$ and on reaching $x=0$, it produces HD profile having a left boundary layer. Further, density profile with a shock also develops a boundary layer at left in lane-$A$ in the region $\alpha > 1/2$ (Fig. \ref{fig4}(d)). Similar kind of transition from shock to HD profile is identified if one moves from (S,HD) phase vertically upwards in $\alpha-\gamma$ phase-plane. The boundaries of (HD,HD) phase are given by $\alpha=1-\rho_{A,o}(\gamma)$ (for $\alpha < 1/2$) and $\gamma=\gamma_{cA}$ (for $\alpha > 1/2$). Fig.~\ref{fig4}(e) shows the system in (HD,HD) phase.

The (LD,HD) phase can be reached either from (LD,S) phase or from (HD,S) phase through vertical transition line $\alpha=\alpha_c$ or horizontal transition line $\gamma=\gamma_c$, respectively. The density profile in (LD,HD) phase are diversified in terms of nature of boundary layers. It is evident from Fig.~\ref{fig1}, where (LD,HD) phase is subdivided into nine subregions by surface transition lines. Each of this subregion identifies a distinct density profile in both the lanes. One of such profiles, having coth-r with left boundary layer in lane-$A$ and coth-l with right boundary layer in lane-$B$, is shown in Fig.~\ref{fig4}(f).

\section{\label{SD} Shock dynamics}
A complete and nice description of formation (merging) of shocks from (into) boundary layers is presented in the form of phase diagram (Fig. \ref{fig3}). Here, we are interested in investigating the dynamics of shock and their variations with respect to the system parameters.
A discontinuity in the bulk connecting a low (high)-density part on the left to a high(low)-density part on the right is known as upward (downward) shock. The nature of shock can be examined with the help of total current in lane-$j$, given by $J_j=-\frac{\epsilon}{2}\frac{d \rho_j}{d x}+ \rho_j(1-\rho_j)$. For an upward (downward) shock i.e. $d\rho_j/dx > 0$ ($d\rho_j/dx < 0$), we have $J_j < \rho_j(1-\rho_j)$ ($J_j > \rho_j(1-\rho_j)$).

\begin{figure}
\includegraphics[height=5cm,width=7cm]{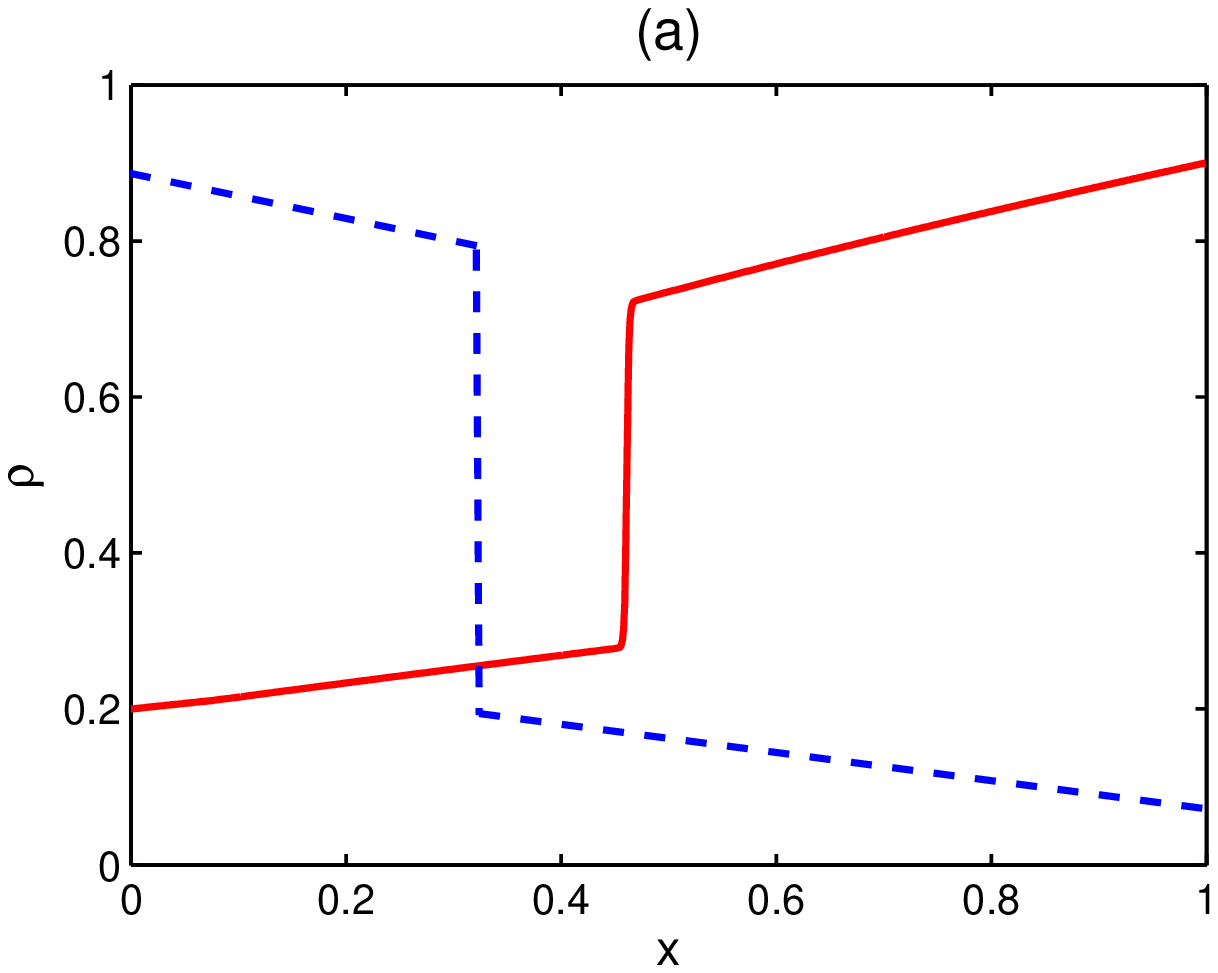}
\includegraphics[height=5cm,width=7cm]{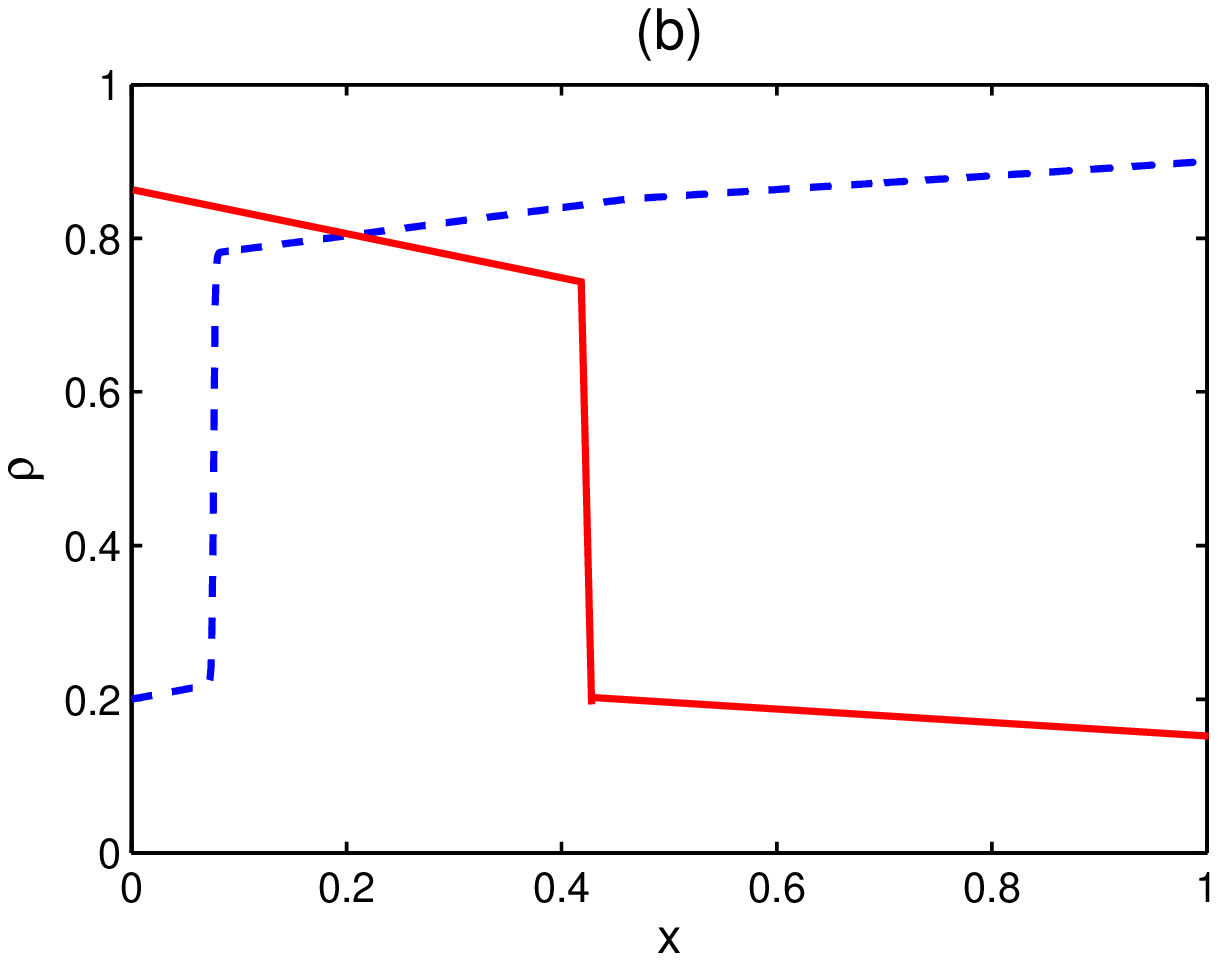}
\caption{\label{fig10} (Color online) The solid (dashed) curves indicate density profiles in lane-$A (B)$.}
\end{figure}

The particular observation which is of interest to us is that there is no possibility of existence of a downward shock in our system. The non-existence of downward shock in lane-$B$ would imply the same for lane-$A$ and vice-versa. This point is validated graphically in Fig. \ref{fig10}. Fig. \ref{fig10}(a) shows the situation when density in lane-$A$ incurs an upward shock and density in lane-$B$ has a downward shock. The reverse situation is shown in Fig. \ref{fig10}(b). Both the situations violate the condition $\rho_A < \rho_B$ (section \ref{PD}). Therefore, it is sufficient to show that there does not exist a downward shock in lane-$B$. This is justified with the help of fixed point diagram ~\cite{mukherji2009fixed, yadav2012phase}. Since, one can ignore the contribution of particle non-conserving terms in the boundary layer or shock regions, we set these terms to zero in the system \eqref{eq15} and \eqref{eq16}. It gives $\rho_A=f(\rho_B)$ (say), where $f$ is a function of $\rho_B$.

Adding Eqs. \eqref{eq15} and \eqref{eq16}, we get
\begin{eqnarray}
\frac{\epsilon}{2}\bigg(\frac{d^2 \rho_A}{d x^2}+\frac{d^2 \rho_B}{d x^2}\bigg)+(2\rho_A-1)\frac{d\rho_A}{dx}+(2\rho_B-1)\frac{d \rho_B}{dx}=0. \label{eq24}
\end{eqnarray}
Integrating Eq. \eqref{eq24} with respect to $x$, we get
\begin{eqnarray}
\frac{\epsilon}{2}\bigg(\frac{d \rho_A}{d x}+\frac{d\rho_B}{d x}\bigg)+ \rho_A^2-\rho_A+\rho_B^2-\rho_B=c. \label{eq25}
\end{eqnarray}
Here, $c$ is a constant of integration. The fixed points of Eq. \eqref{eq25} are given by $\rho_A^2-\rho_A+\rho_B^2-\rho_B=c$. Substituting $\rho_A=f(\rho_B)$, we obtain two-dimensional fixed point diagram in $c-\rho_B$ plane (Fig. \ref{fig10}).

The fixed point diagram contains two branches $ab$ and $bc$. The stability analysis shows that the lower branch ($ab$) is unstable while upper branch ($bc$) is stable. Geometrically, a downward shock is possible if a point on the curve in upper branch can be connected to a point in the lower branch by a vertical line \cite{yadav2012phase}. One can easily see from the direction of vertical arrows that it is not possible to get a downward shock in lane-$B$. So, we can not get downward shock in the bulk of density profiles in any of the two lanes. It is also clear from the density profiles in various phases obtained with the continuum mean-field results as well as Monte-Carlo simulation results (Fig. \ref{fig4}). This is an important result which supports our previous observation that a decaying boundary layer at either boundary can not produce a shock through its deconfinement.
\begin{figure}[h]
\begin{center}
\includegraphics[height=6cm,width=8cm]{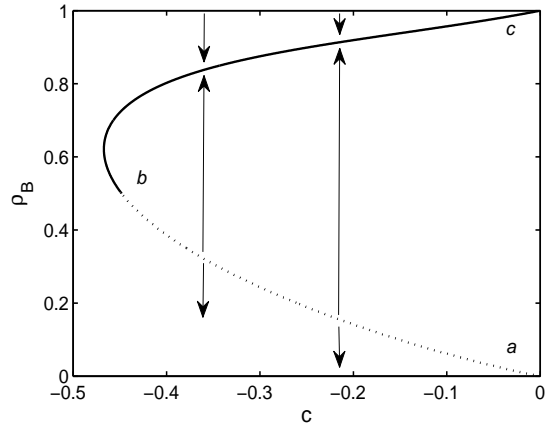}
\caption{\label{fig11} Fixed point diagram for two-channel system. The branches $ab$ and $bc$ have been referred in text as lower and upper fixed point branches. Vertical arrows show the stability nature of two branches.}
\end{center}
\end{figure}

The shock (domain wall) in the bulk is characterized by two important quantities: position and height. We denote the position of domain wall in lane-$j$ by $x_{s,j}$. Since, particle non-conserving dynamics do not play any role in the shock region, we find $x_{s,j}$ using the constancy of current across the shock given by
\begin{equation}
\rho_{j,+}^2-\rho_{j,+}=\rho_{j,-}^2-\rho_{j,-}. \label{eq26}
\end{equation}
Here, $\rho_{j,-}=\displaystyle\lim_{x\rightarrow x_{s,j}^{-}}\rho_j(x)$ and $\rho_{j,+}=\displaystyle\lim_{x\rightarrow x_{s,j}^{+}}\rho_j(x)$.
The condition given by Eq. \eqref{eq26} is equivalent to $\rho_{j,+}+\rho_{j,-}=1$, which has been used to capture the shock positions.

\begin{figure}[h]
\includegraphics[height=5cm,width=7cm]{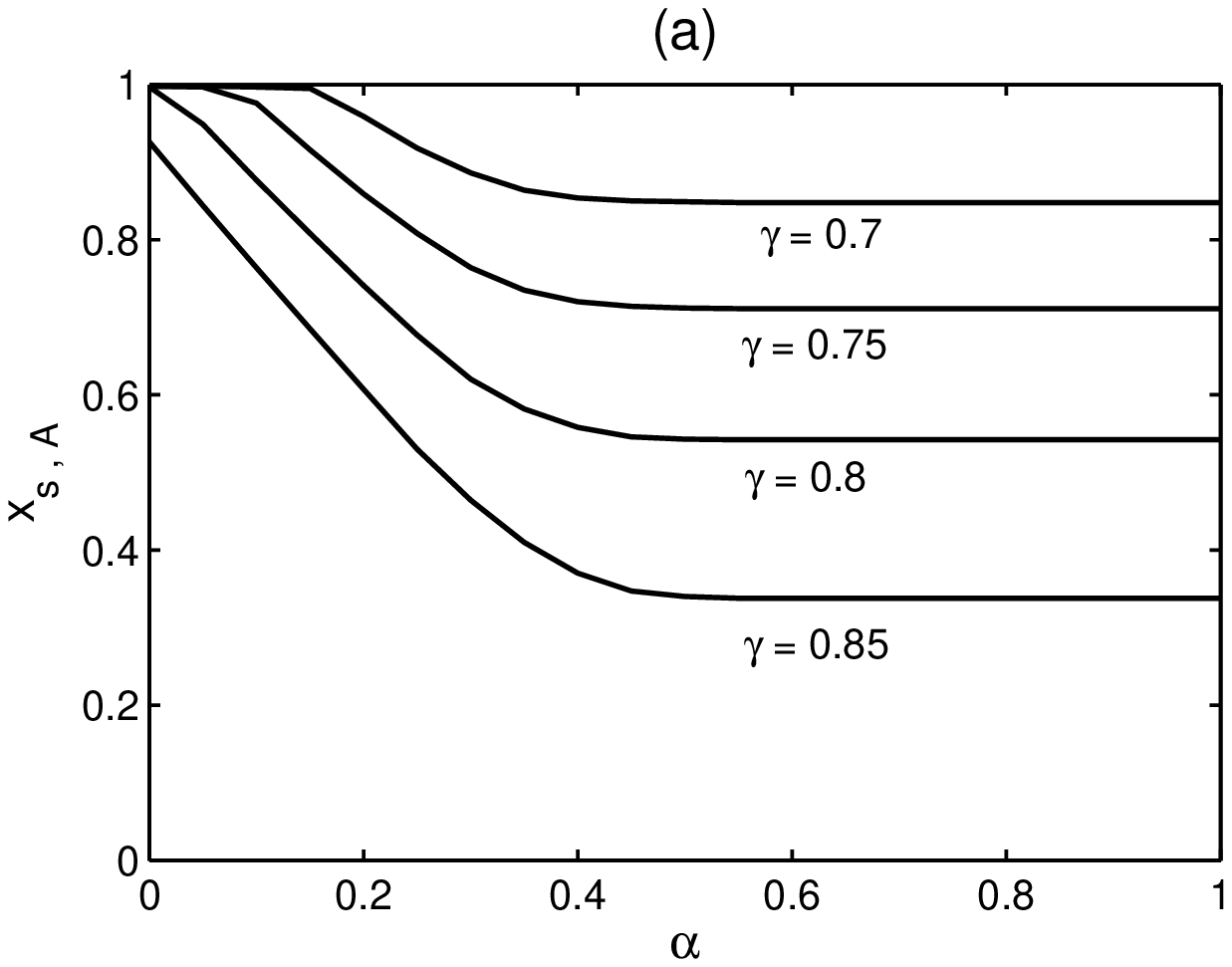}
\includegraphics[height=5cm,width=7cm]{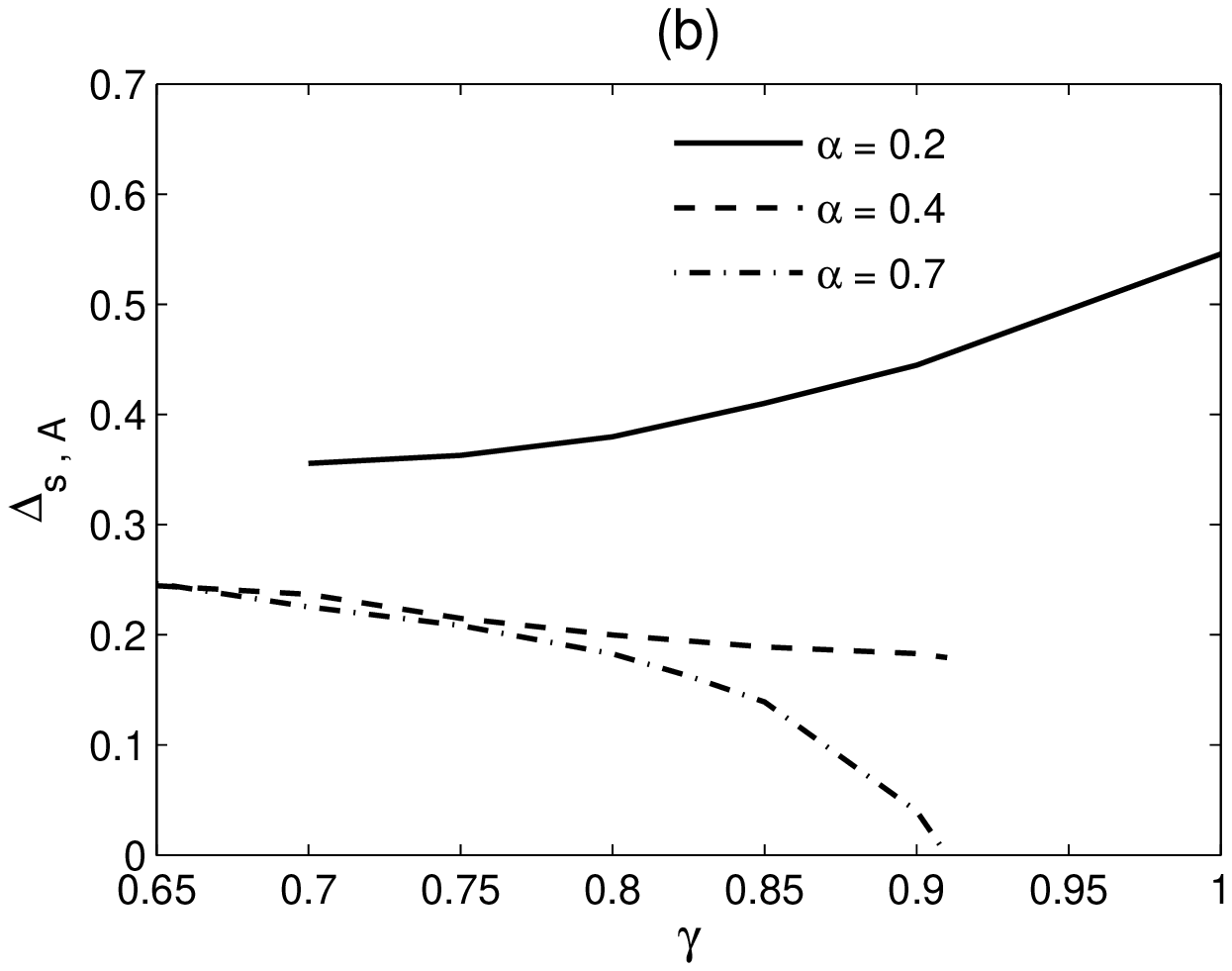}
\caption{\label{fig12}The dependence of (a) The location of domain wall on $\alpha$ for different values of $\gamma$. (b) Height of shock on $\gamma$ for different values $\alpha$. }
\end{figure}

Fig. \ref{fig12} shows the variations of position of domain wall with respect to $\alpha$ for different values of $\gamma$ in lane-$A$. Upon increasing $\alpha$ and hence inflow of particles, there is a continuous (linear) change in the position of domain wall from right to left in the bulk till $\alpha \leq 1/2$. A gradual settlement of the domain wall to a fixed position in the bulk is observed when $\alpha > 1/2$. The increase in inflow of particles enlarges the HD part in the density profile and hence domain wall moves leftwards. When $\alpha \geq 1/2$, there appears a boundary layer at the entrance which obstructs the incoming particles from entering the bulk and hence the domain wall in the bulk does not move further. This leads to the localization of shock with respect to the entrance rate $\alpha$.
Fig. \ref{fig12}(b) shows the variation of height of domain wall in lane-$A$ ($\Delta_{s,A}=\rho_{A,+}-\rho_{A,-}=2\rho_{A,+}-1$) with respect to $\gamma$ along lines of constant entrance rate $\alpha$. For $1-\rho_{A,o}(\gamma) < \alpha < 1/2$, $\Delta_{s,A}$ jumps discontinuously to a finite value on entering shock region from HD phase, whereas there is a continuous rise in the height from zero for $\alpha \geq 1/2$. Further, at the phase boundary between LD and shock phase ($ \gamma = 1 - \rho_{A,o}(\alpha)$), $\Delta_{s,A}$ jumps to zero discontinuously.
\section{\label{effect}Effect of coupling strength on steady-state dynamics}
\begin{figure*}[htb]
\begin{center}
\includegraphics[height=4cm,width=5cm]{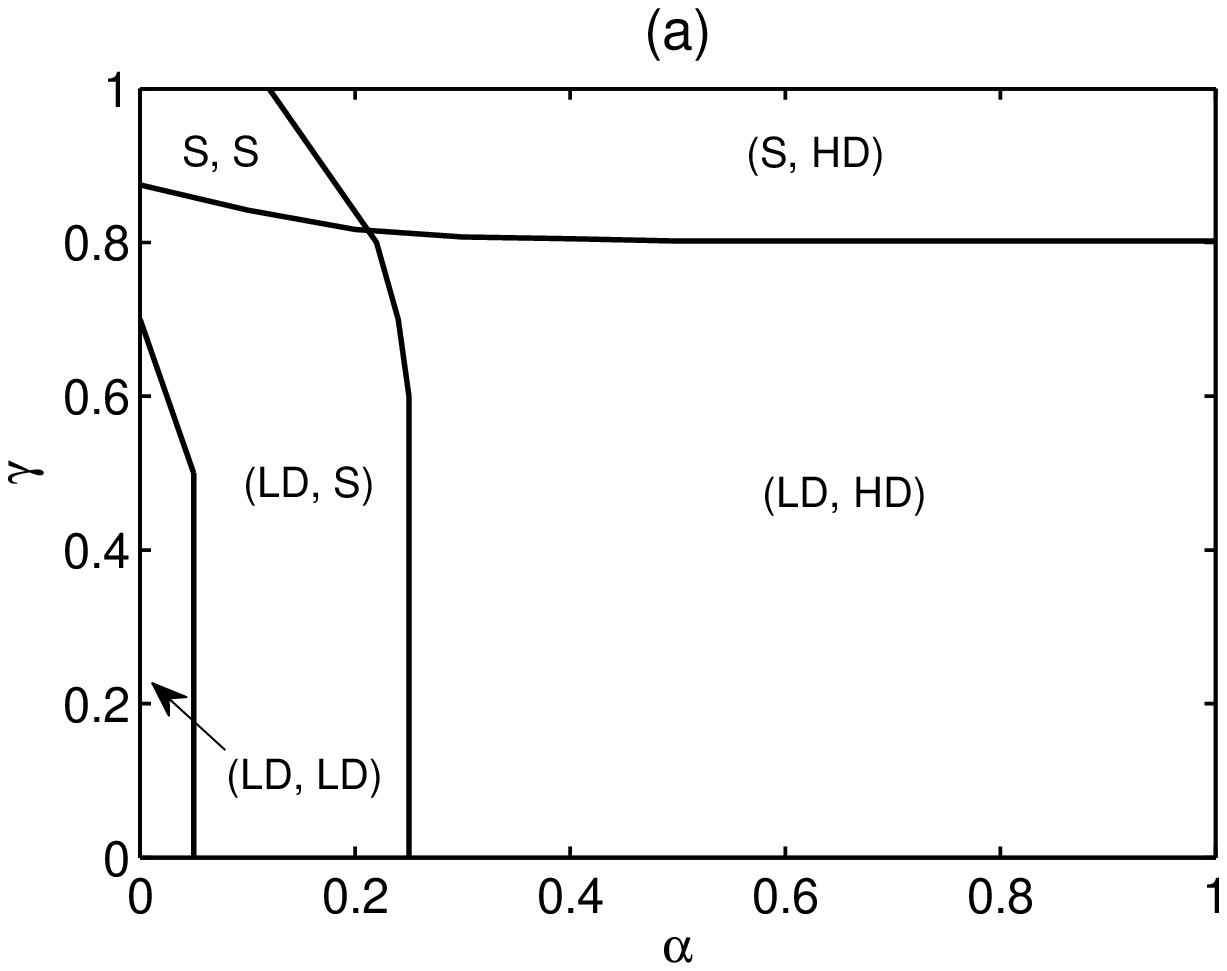}
\includegraphics[height=4cm,width=5cm]{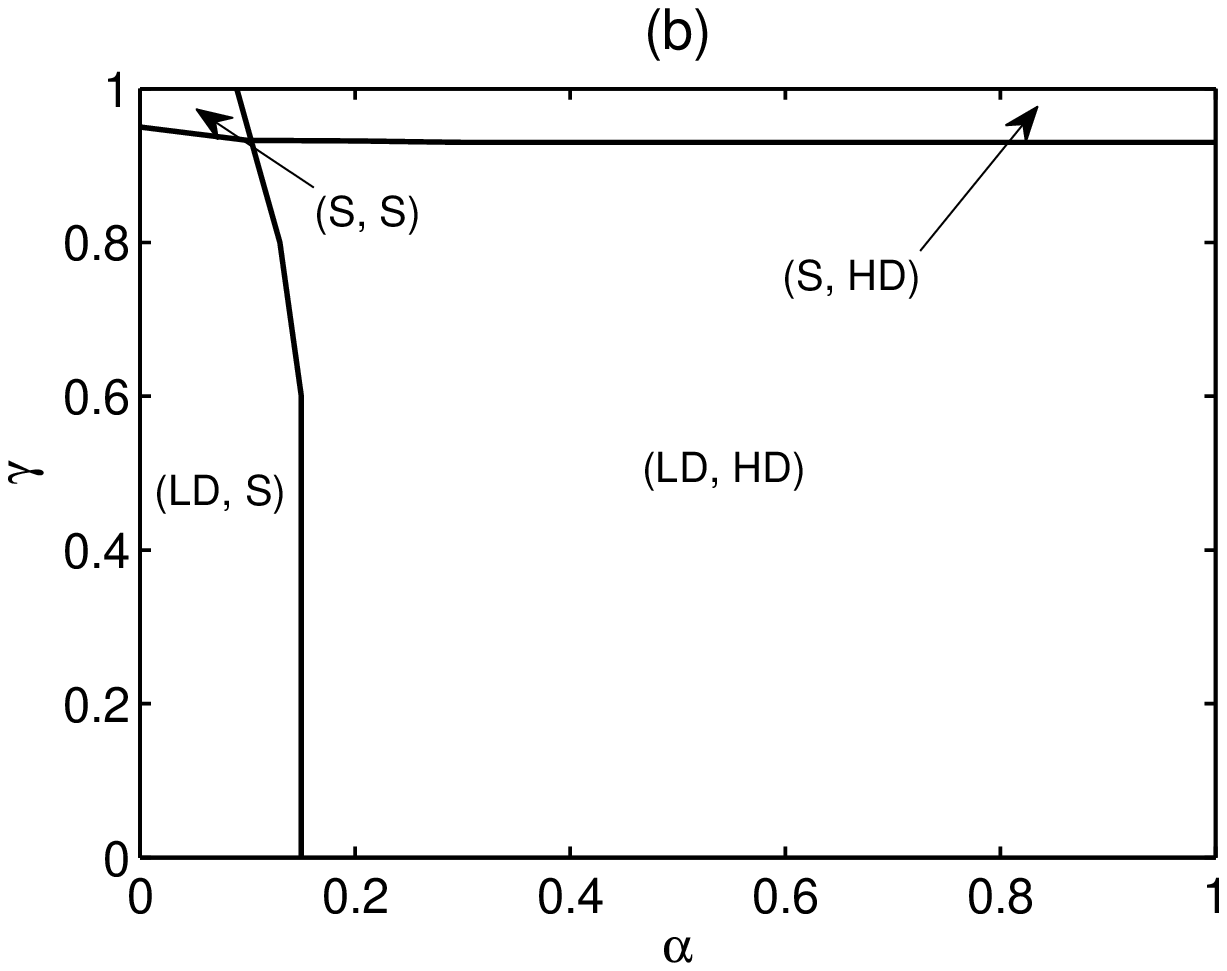}
\includegraphics[height=4cm,width=5cm]{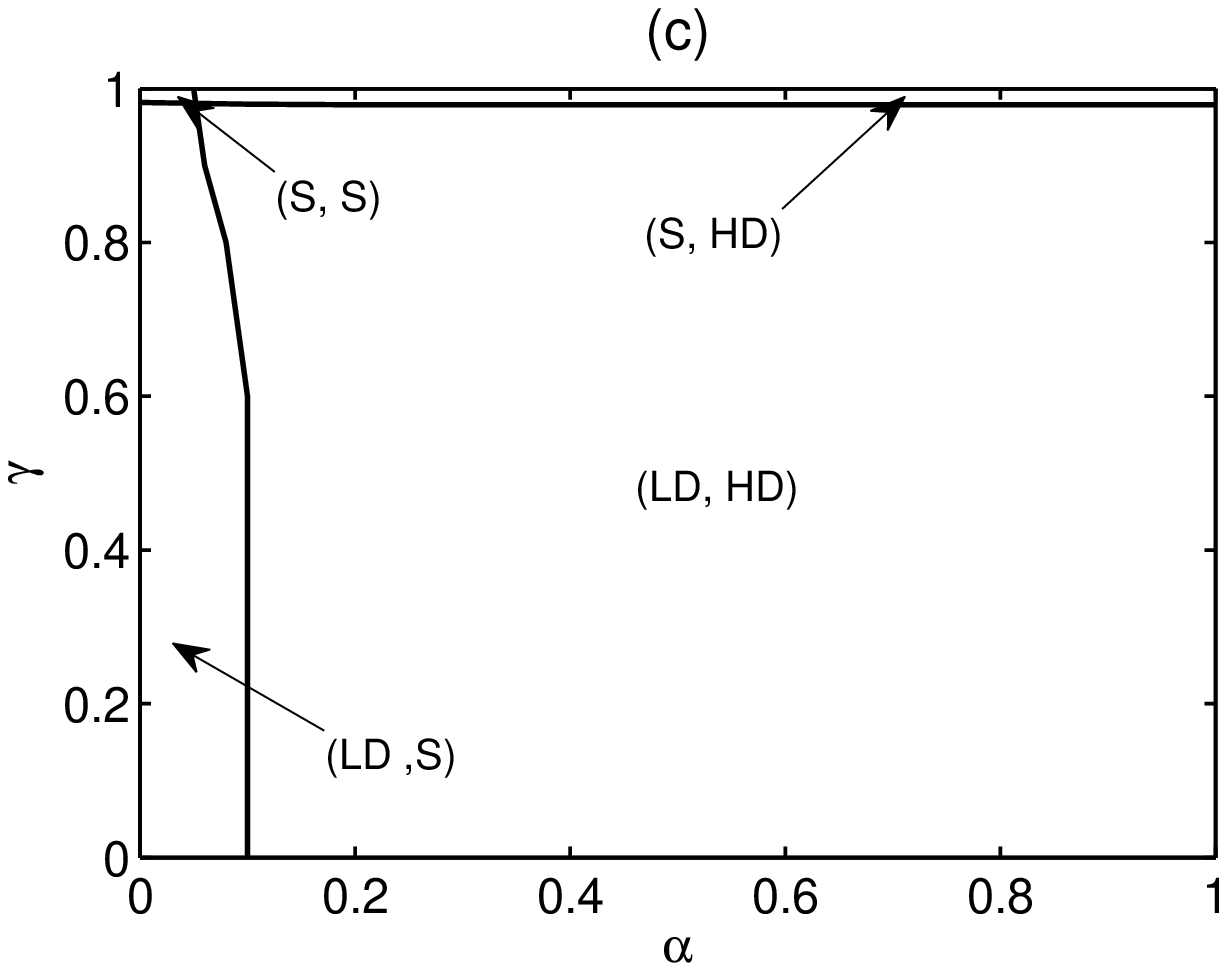}
\caption{\label{fig13}
 The phase diagrams with $\Omega_d=0.2$ and (a) $\Omega=10$, (b) $\Omega=100$ and (c) $\Omega=1000$}
\end{center}
\end{figure*}

So far, we have explored the system dynamics for small value of lane-changing rate i.e. $\Omega=1$. Now, we wish to examine the effect of increasing the coupling strength on the phase diagram of the system. Fig. \ref{fig13}(a) shows the steady-state phase diagram for our two-channel system with $\Omega=10$ keeping all other parameters fixed. Physically, the increase in lane-changing rate increases the number of vertical transitions from lane-$A$ to $B$, which eventually decreases (increases) the particle density in lane-$A(B)$. As a result, in $\alpha-\gamma$ phase-plane, the region confined to LD phase enlarges and that to HD phase contracts. The reverse phenomenon happens for phases in lane-$B$. This combined effect can be seen from the phase diagram, where the major part of $\alpha-\gamma$ phase-plane is covered by (LD,HD) phase. Here, it is important to note that the number of steady-state phases is reduced to five due to the disappearance of (HD,HD) phase. The particle density in lane-$A$ is not enough to produce a HD profile and hence there cannot exist (HD,HD) phase in the steady-state of the system.
A qualitatively similar effect is observed on further increasing the coupling strength. For $\Omega=100$, the number of steady-state phases reduces to four with the exclusion of (LD,LD) phase from the $\alpha-\gamma$ phase-plane. Additionally, the area covered by (LD,HD) phase expands further. When the lane-changing rate is increased to its maximum value i.e. $\Omega=1000$, one finds that (S,S), (LD,S) and (S,HD) phases are confined to a very small region near the boundaries of the phase-plane.
\section{\label{conclusion}Conclusion}
In this work, we have studied a two-lane totally asymmetric simple exclusion process with open boundaries. The particles in the bulk follow attachment-detachment kinetics from a bulk reservoir with certain rates ($\omega_a, \omega_d$). Additionally, particles in lane-$A$ can shift to lane-$B$ through a vertical transition with rate $\omega$.

A detailed study of the steady-state properties of the system has been carried out using boundary layer analysis of mean-field equations in the continuum limit under the particular case of $\omega_a=\omega_d$. A qualitative comparison of the phase diagram of our asymmetrically coupled two-lane system with that of a symmetrically coupled two-lane TASEP with LK reveals that the structure of the former is quite complex as compared to the latter. The lines of phase transitions have been quantified in terms of boundary rates. It has been found that the phase transition from LD (HD) to shock phase occurs through deconfinement of right (left) boundary layer. Another kind of transition is the surface transition occurring in both LD and HD phases which is associated with the divergence of a length-scale $\xi_j$. This transition does not correspond to a phase change; rather it changes the sign of slope of the boundary layer.

An important result about the non-existence of a downward shock in the system has been analyzed with the help of fixed-point theory. Along with the formation of shock, we have investigated various other characteristics of shock such as its motion in the bulk, position and height. The dependence of shock dynamics on boundary rates has been examined. Moreover, the effect of lane-changing rate on the steady-state dynamics of the system has been studied. A significant effect of coupling strength is the reduction in number of steady-state phases in the system, which reduces to five and then to four gradually with increasing lane-changing rate.

In this paper, we have analyzed the steady-state properties of the proposed model for the particular case of equal attachment and detachment rates. Our approach can be generalized to investigate the system in various other cases such as unequal attachment and detachment rates and to study the asymmetric coupling conditions in which particles can move from any one lane to another with unequal rates. The present study might help not only in understanding complex dynamics of motor proteins but also towards enhancement of one's insight in non-equilibrium statistical mechanics.

\begin{acknowledgments}
The second author acknowledges CSIR, New Delhi, India, for providing financial support.
\end{acknowledgments}

\end{document}